\documentclass[twocolumn,superscriptaddress,amsmath,amssymb,aps,prr]{revtex4-1}
\usepackage{graphicx}
\usepackage{amssymb}
\usepackage{amsmath}
\usepackage{mathrsfs}

\usepackage{multirow}
\usepackage{array,tabularx}
\usepackage{xcolor}
\usepackage{comment}
\usepackage{dcolumn}
\usepackage{bm}

\usepackage[colorlinks,urlcolor=blue,citecolor=blue,linkcolor=blue]{hyperref}

\newcommand{\ve}[1]{{\mathbf #1}}

\newcommand{\Frac}[2]{\displaystyle\frac{#1}{#2}}

\renewcommand{\k}{{\bf k}}

\newcommand{\q}{{\bf q}}
\newcommand{\Q}{{\bf Q}}
\newcommand{\0}{{\bf 0}}
\newcommand{\sch}{Schr{\"o}dinger }

\newcommand{\area}{\mathcal{A}}
\newcommand{\ef}{E_{F}}
\newcommand{\kF}{k_{F}}
\newcommand{\eb}{\varepsilon_B}
\newcommand{\rvec}{\mathbf{r}}
\newcommand{\Rvec}{\mathbf{R}}

\newcommand{\bra}[1]{\langle{#1}|}
\newcommand{\ket}[1]{|{#1}\rangle}

\usepackage{float}
\usepackage{upgreek}
\usepackage{makeidx}
\usepackage{enumerate}
\DeclareGraphicsExtensions{.png,.jpg,.pdf,.mps,.gif,.bmp,.eps}
\graphicspath{{figures/}}
\usepackage{physics}

\begin{document}



\title{Effect of fermion indistinguishability on optical absorption of doped two-dimensional 
semiconductors}


\author{A. Tiene}
\email{antonio.tiene@uam.es}
\affiliation{Departamento de F\'isica Te\'orica de la Materia
  Condensada \& Condensed Matter Physics Center (IFIMAC), Universidad
  Aut\'onoma de Madrid, Madrid 28049, Spain}

\author{J. Levinsen}
\affiliation{School of Physics and Astronomy, Monash University, Victoria 3800, Australia}
\affiliation{ARC Centre of Excellence in Future Low-Energy Electronics Technologies, Monash University, Victoria 3800, Australia}

\author{J. Keeling}
\affiliation{SUPA, School of Physics and Astronomy, University of St Andrews, St Andrews, KY16 9SS, United Kingdom}

\author{M. M.~Parish}
\affiliation{School of Physics and Astronomy, Monash University, Victoria 3800, Australia}
\affiliation{ARC Centre of Excellence in Future Low-Energy Electronics Technologies, Monash University, Victoria 3800, Australia}

\author{F. M. Marchetti}
\email{francesca.marchetti@uam.es}
\affiliation{Departamento de F\'isica Te\'orica de la Materia
  Condensada \& Condensed Matter Physics Center (IFIMAC), Universidad
  Aut\'onoma de Madrid, Madrid 28049, Spain}

\date{\today}

\begin{abstract}
 We study the optical absorption spectrum of a doped two-dimensional semiconductor in the spin-valley polarized limit. In this configuration, the carriers in the Fermi sea are indistinguishable from one of the two carriers forming the exciton. Most notably, this indistinguishability requires the three-body trion state to have $p$-wave symmetry. To explore the consequences of this, we evaluate the system's optical properties within a polaron description, which can interpolate from the low density limit---where the relevant excitations are few-body bound states---to higher density many-body states.
 In the parameter regime where the trion is bound, we demonstrate that the spectrum is characterized by an attractive quasiparticle branch, a repulsive branch, and a many-body continuum, 
 and we evaluate the doping dependence of the corresponding energies and spectral weights.
 In particular, at low doping we find that the oscillator strength of the attractive branch scales with the square of the Fermi energy as a result of the trion's $p$-wave symmetry. Upon increasing density, we find that both the repulsive and attractive branches blueshift, 
 and that the orbital character of the states associated with these branches interchanges. 
We compare our results with
previous investigations of the scenario where the Fermi sea involves carriers distinguishable from those in the exciton,  for which the trion ground state is $s$-wave.
\end{abstract}


\maketitle

\section{Introduction}

Trions---the bound state of either two conduction-band electrons and a
valence-band hole ($X^{-}$) or one electron and two holes ($X^{+}$)---can be thought of as the semiconductor analog of the hydrogen anion $H^{-}$ and molecular ion $H_2^{+}$, respectively~\cite{Lampert_PRL1958}. In practice,
trions  are observed in electronically doped semiconductors, with a finite excess density of electrons (or holes) coexisting with the $X^-$ (or $X^+$) state.
While the trion description in terms of an isolated three-body state may apply in the limit of a very low density of excess majority particles~\cite{Glazov_JCP2020}, a more appropriate picture that spans the whole range from low to higher densities is that of polarons~\cite{Efimkin_PRB2017,Sidler_NP2016,Efimkin_PR2021,Rana_PRL2021}.
%
Here, an exciton is dressed by the Fermi sea of excess charge carriers, forming repulsive and attractive exciton-polaron branches. This leads to absorption peaks that, in the limit of vanishing doping, connect smoothly to the underlying exciton and trion few-body states. Exciton-polarons have recently generated significant interest since they have been used to interpret the absorption peaks of charge-doped two-dimensional (2D) semiconductors~\cite{Sidler_NP2016,Tan_PRX2020,Emmanuele_NC2020,Ke_JPCL2021,Koksal_arxiv2021}.

The physics of repulsive and attractive polarons---and their connection to few-body bound states---has been extensively studied in the context of polarized fermionic atomic gases~\cite{Schirotzek2009,Nascimbene2009,Kohstall2012,Koschorreck2012,Zhang2012,Wenz2013,Cetina2015,Ong2015,Cetina2016,Scazza2017,Yan2019,Oppong2019,Ness2020}---
see the recent reviews of such atomic polarons in two~\cite{2Dreview} and three dimensions~\cite{Massignan_Zaccanti_Bruun}.  However, new questions arise when considering such states in semiconductors.  In particular, the natural experimental probe of exciton-polarons in semiconductors is optical absorption,  which necessarily probes transitions to states exciting inter-band particle-hole pairs, i.e., an even number of fermions.
The significant difference compared to atomic polaron physics is that when we create an exciton, both the electron and the hole interact with the Fermi sea.  By contrast, for atoms, the typical experiment is injection RF spectroscopy, which involves flipping an impurity atom's state from non-interacting to interacting with the fermionic medium~\cite{Punk2007,Liu2020}.  This process can also be thought of as the creation of a particle-hole pair: the particle is the atom in the interacting state, the hole is the missing atom in the non-interacting state.  As  such in this case the ``hole'' does not interact with the Fermi sea (nor does the hole interact with the flipped atomic state).

\begin{figure}
\includegraphics[width=0.45\textwidth]{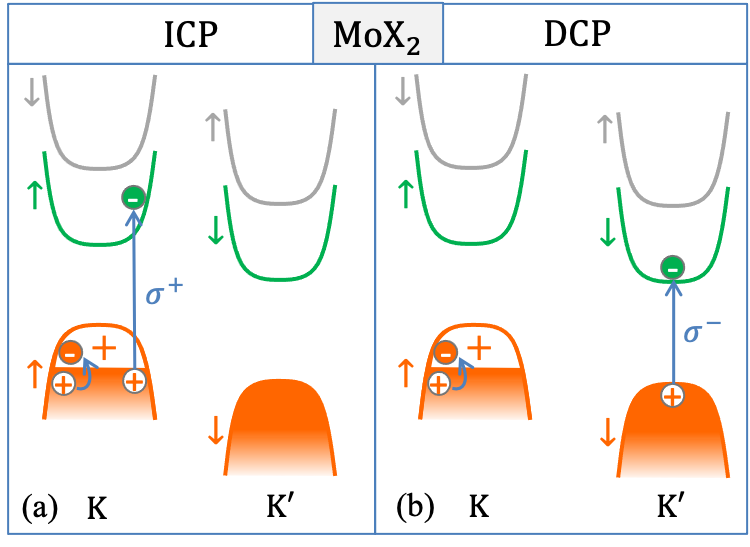}
\caption{Illustration of the two classes of polaron dressing. The Fermi sea of holes is either (a) indistinguishable or (b) distinguishable from the hole forming the exciton, leading to the indistinguishable (ICP) and distinguishable carrier polaron (DCP) scenarios, respectively.
These two different cases can arise when spin-valley Zeeman splitting is induced via an external out-of-plane magnetic field, combined with using circularly polarized light to selectively create excitons either in the same (a) or the opposite (b) valley as the Fermi sea.}
\label{fig:schematic}
\end{figure}

In the semiconductor case, the Fermi sea can involve carriers that are either indistinguishable or distinguishable from those contained in the exciton that forms the exciton-polaron. For brevity, we will refer to these two cases as the \emph{indistinguishable} (ICP) and the \emph{distinguishable carrier polaron} (DCP) scenarios. Figure~\ref{fig:schematic} illustrates these two cases for the hole-doped $X^+$ case. In the special case of degenerate conduction bands, the exciton may also be dressed by both Fermi seas (not shown). 
It is generally accepted that the absorption spectrum observed experimentally in Ref.~\cite{Sidler_NP2016} arises primarily from the dressing of the exciton by a distinguishable Fermi sea---the DCP case, as shown in Fig.~\ref{fig:schematic}(b).  As such, that scenario has received the most theoretical attention~\cite{Efimkin_PRB2017,Sidler_NP2016,Rana_PRL2021,Efimkin_PR2021}. As we will discuss in this paper, the ICP case 
shown in Fig.~\ref{fig:schematic}(a), 
leads to qualitatively different features in the optical absorption spectra and its evolution with doping.

The key difference between the ICP and DCP scenarios is the requirement of overall antisymmetry for fermionic wave functions. As a consequence,
when the trion contains two indistinguishable excess majority particles---identical spin and valley indices---the lowest energy trion state has one unit of angular momentum, i.e., it is a $p$-wave state. By contrast, for distinguishable particles, the ground state is $s$-wave.
$X^{-}$ trions with distinguishable electrons---$s$-wave trions---have been studied in  both II-VI and III-V quantum wells~\cite{Kheng_PRL1993,Finkelstein_PRL1995,Huard_PRL2000}, as well as in transition metal dichalcogenides (TMD) monolayers~\cite{Ross_NatComm2013,Mak_NatMat2013,Jones_NP2016,Plechinger_NatCom2016,Vaclavkova_Nano2018}.
In contrast to $s$-wave trions, observations of $p$-wave trions have been scarce, since for many materials this state is unbound.
In general, it is known that, for the positively charged trions $X^{+}$, binding can occur when the ratio of electron to hole mass is sufficiently small~\cite{Courtade_PRB2017}.
Indeed, for TMD monolayers, calculations for MoX$_2$ (X$=$Se, S) indicate that the intravalley $p$-wave trion is unbound~\cite{Tempelaar_NC2019}, which is consistent with the effective masses of electrons and holes being too similar to permit a $p$-wave bound state.
A similar statement holds also in CdTe- and GaAs-based quantum wells~\cite{Sergeev_PSS2001,Sergeev_Nanot2001}.
The $p$-wave trion has, however, been predicted to exist in the presence of a magnetic field perpendicular to the quantum well~\cite{Sandler1992,Macdonald1992two,Dzyubenko1994,Palacios_PRB1996,whittaker_PR1997}, and this has been observed experimentally~\cite{Shields_PRB1995,Finkelstein_PRB1996,Sanvitto_PRL2002}.
There has also been evidence of $p$-wave trions in an electric field \cite{Shields_PRB1997}.

In this paper, we explore how the optical spectrum of the ICP case evolves with doping, and identify how this differs from the DCP case. 
At low doping, the $p$-wave trion  state has a vanishing oscillator strength for two reasons: the low probability of finding a nearby excess charge (as also applies to the $s$-wave trion~\cite{Shiau_EPL2017,Glazov_JCP2020,Zhumagulov_PRB2020,zhumagulov2021microscopic,Rana_PRL2021}), 
and the vanishing dipole matrix element of the $p$-wave trion.
Preliminary investigations have suggested that, in the presence of a magnetic field (required for binding), the $p$-wave trion acquires an oscillator strength which increases with doping~\cite{Sanvitto_PRL2002}.
We will illustrate the mechanism behind the increase of the oscillator strength with doping
and show that, for $p$-wave trions, the oscillator strength grows like the square of the Fermi energy and, thus, is slower than for $s$-wave trions where it scales linearly with Fermi energy.

While the few-body description is relevant at low doping, optical absorption in doped 2D semiconductors is better described by the exciton-polaron scenario, where the exciton is dressed by 
an excited intra-band particle-hole pair of the majority Fermi sea  (see Fig.~\ref{fig:schematic}).
For the ICP case in the low doping limit, the properties of the  attractive (repulsive) polaron branch recover those of the trion (exciton). However, when doping increases, we observe an evolution of the nature of the two branches which goes beyond few-body physics. In particular, by evaluating the angular momentum of the majority-species hole within the dressing cloud of the attractive polaron quasiparticle, we observe that the hole changes from angular momentum $\pm 1$ ($p$-wave symmetry) at low doping to angular momentum $0$ ($s$-wave) at large doping.
The repulsive branch instead has the opposite evolution with increasing doping, switching over from an $s$-wave symmetry to a $p$-wave symmetry.

As the character of the attractive polaron quasiparticle changes with increasing doping, we also find that there is a significant transfer of oscillator strength to an incoherent many-body continuum, which becomes the dominant contribution at large doping. Yet, we find that the effect of this continuum on the linewidth of the repulsive polaron branch is negligible, due to a difference in symmetry of the repulsive branch and the continuum in this regime. This behavior is qualitatively different from the DCP case: In that case, the oscillator strength of the continuum is strongly suppressed, while the coupling between the repulsive branch and the continuum is allowed by symmetry such that the repulsive polaron linewidth grows with doping~\cite{2Dreview,Sidler_NP2016,Efimkin_PR2021}.

The oscillator strengths and linewidths of these different branches have consequences for the polariton spectrum found when there is strong matter-light coupling~\cite{Sidler_NP2016,Tan_PRX2020,Emmanuele_NC2020,Ke_JPCL2021,Koksal_arxiv2021}. We show how the system's spectral properties in the weak-coupling limit determine the polariton spectrum in strong coupling.  Specifically, we demonstrate that increasing the electron doping leads to the appearance of three polariton branches, corresponding to strong coupling with both attractive and repulsive polarons.

The paper is organized as follows: In Sec.~\ref{sec:model}, 
we introduce the model describing an interacting electron-hole system coupled to light, where both electrons and holes are indistinguishable. 
In Sec.~\ref{sec:spin-pwave-t}, we provide a summary of the properties of $p$-wave trion states when the interaction between charges are strongly screened.
In Sec.~\ref{sec:M4}, we derive the properties of an exciton-polaron (or exciton-polariton-polaron) state, describing excitons (or polaritons) in a   charged imbalanced system with indistinguishable excess majority particles. By evaluating the system spectral function in Sec.~\ref{sec:GreenF}, we are able to describe the doping mediated transfer of oscillator strength from the repulsive branch to the attractive branch,  as well as the crossover of their symmetry properties. 
Conclusions and perspectives are gathered in Sec.~\ref{sec:Conc}.

\section{Model}
\label{sec:model}
%
\subsection{Model Hamiltonian}
\label{sec:modelH}
We seek to model the spectral response of a doped semiconductor,
where one of the two charges forming the exciton is identical to those forming the Fermi sea.  As such, we write the following Hamiltonian, describing only two species of charges, each belonging to a single (spin-polarized) conduction and valence band 
(throughout this paper we work in units where $\hbar=1$):
\begin{subequations}
\label{eq:Hamiltonian}
\begin{align}
\label{eq:Hamiltonian_terms}
  \hat{H} &= \hat{H}_0 + \hat{H}_{eh} +
  \hat{H}_{ehC}\\
\label{eq:H_0}
  \hat{H}_0 &=\sum_{\k\sigma} \epsilon_{\k,\sigma}  \hat{c}^\dag_{\k,\sigma} \hat{c}^{}_{\k,\sigma}
  + \sum_{\q} \nu_{\q} \hat{a}_{\q}^{\dag} \hat{a}_{\q}^{} \\
\label{eq:H_eh}
  \hat{H}_{eh} &= -\frac{v}{\area}\sum_{\ve{k}\ve{k'}\ve{q}}
  \hat{c}^\dag_{\ve{k},1} \hat{c}^\dag_{\ve{k'},2}
  \hat{c}^{}_{\ve{k'}+\ve{q},2}
  \hat{c}^{}_{\ve{k}-\ve{q},1}\\
  \hat{H}_{ehC} &= \frac{g}{\sqrt{\area}} \sum_{\k\q}
  \left(\hat{c}^\dag_{\frac{\q}2+\k,1} \hat{c}^\dag_{\frac{\q}2 -
    \k,2} \hat{a}_{\q}^{} + \text{h.c.}\right) \; .
\label{eq:H_ehC}
\end{align}
\end{subequations}
Here, $\hat{c}^{}_{\k,\sigma=1,2}$ and
$\hat{c}^{\dag}_{\k,\sigma=1,2}$ are majority ($\sigma=1$) and
minority ($\sigma=2$) species annihilation and creation operators, respectively. 
These
have a dispersion $\epsilon_{\k,\sigma} = \k^2/2m_{\sigma}$, where
$m_\sigma$ is the effective mass and $\k$ is the two-dimensional (2D)
momentum. Note that, throughout the paper, energies are measured with respect to the 
band gap
 We denote the density of the majority species as $n_1$  and thus the Fermi
energy is,
\begin{equation}
 \ef=\Frac{k_F^2}{2m_1} = \Frac{2\pi}{m_1} n_1\; , 
\label{eq:Fermi_energy}
\end{equation}
where $k_F$ is the Fermi momentum. 
For electron doping, $\sigma=1$ are
conduction electrons and $\sigma=2$ are valence holes.  For
hole doping, $\sigma=1$ are valence holes and $\sigma=2$ are
conduction electrons. 
The only distinction between these two cases is the assignment of masses $m_{1,2}$, so we can swap between the two by interchanging $m_1 \leftrightarrow m_2$.

In order to considerably simplify our calculations, in $\hat{H}_{eh}$ we approximate the electron-hole Coulomb interaction as a contact interaction of strength $v>0$  (the factor $\area$ is the system area).
This limit describes the case where interactions between charges are strongly screened. 
In such a case, because of the Pauli exclusion principle, intraspecies interactions vanish. 
One may wonder whether the use of contact interactions---in place of Coulomb, or screened Coulomb interactions---significantly changes when a $p$-wave trion state exists.  
However, as we show in Sec.~\ref{sec:spin-pwave-t}, in the zero-density limit, our model predicts nearly the same critical mass ratio for trion formation~\cite{Pricoupenko_PRA2010,Parish_PRA2013}  as found for the Coulomb problem~\cite{Sergeev_PSS2001,Sergeev_Nanot2001}.
As such, we expect that the use of contact interactions will allow us to faithfully capture the qualitative features of the ICP scenario.

The operators $\hat{a}_{\q}^{}$ and $\hat{a}_{\q}^{\dag}$ describe the cavity photon mode with a
dispersion $\nu_{\q} = \nu_{\0}+\q^2/2m_{C}$, where $m_C$ is the photon mass and $\q$ is the in-plane momentum. These photons couple to the matter excitations via the term $\hat{H}_{ehC}$ in Eq.~\eqref{eq:H_ehC}. We have taken the strength $g$ of the coupling to be independent of momentum and applied the rotating wave approximation. These approximations are appropriate when the band gap greatly exceeds the other energy scales in the problem.

The model we write could be considered as, for example, describing a TMD monolayer with full spin-valley polarized doping, and where the use of circular polarized light allows one to selectively create excitons in the same valley as the Fermi sea. 
Such valley degeneracy breaking---induced by Zeeman splitting via an external out-of-plane magnetic field---has been recently demonstrated in ${\mathrm{WSe}}_{2}$~\cite{Srivastava_NatPhys2015,Aivazian_NatPhys2015}
and in ${\mathrm{MoSe}}_{2}$~\cite{Li_PRL2014,MacNeill_PRL2015}. Indeed, a high degree of valley polarization has been realized at modest magnetic fields, up to an electron density $n_1\simeq 1.6 \times 10^{12}$~cm$^{-2}$ (corresponding to a Fermi energy $\ef \simeq 15$~meV)~\cite{Back_PRL_2017}. 
Alternatively, large valley splittings have also been achieved using the interfacial magnetic exchange field effect~\cite{Zhao_NatureNanot2017}.
The smaller $g$-factors in III-V and II-VI quantum wells mean that it is hard to attain a spin-polarized regime without also realizing Landau quantization of the electron motion.

\subsection{Renormalization of contact interactions}
\label{sec:renormalization}
The use of contact interactions and momentum-independent light-matter coupling introduces ultra-violet (UV) divergences (see, e.g., Ref~\cite{Lawrence1991}).
These can be regularized by introducing an UV cut-off $\Lambda$, i.e., assuming that $v$ and $g$ are non-zero only up to a momentum $\Lambda$, which is typically set by the electronic band structure. 
Recent calculations employing this same model to study polariton-electron scattering~\cite{LiPRL2021,LiPRB2021} showed that results independent of the short-distance physics can be
obtained by then renormalizing both coupling strengths $v$ and $g$ so that observable quantities do not depend on the cut-off. 

The electron-hole interaction strength $v$ can be
renormalized by relating it to the physically measurable exciton binding energy, $\eb>0$, via~\cite{2Dreview}:
\begin{equation}
    \Frac{1}{v} = \Frac{1}{\area} \sum_\k^{\Lambda}
    \frac{1}{\eb +\bar{\epsilon}_{\k}} \; ,
\label{eq:v_renormalization}
\end{equation}
where $\bar{\epsilon}_{\k}=\epsilon_{\k,1}+\epsilon_{\k,2} =
\k^2/2\mu$, with $\mu=(1/m_1+1/m_2)^{-1}$ being the reduced mass. When
$\Lambda \to \infty$, $v^{-1} \sim \ln \Lambda \to \infty$ and thus $v
\to 0$.
Note that the right hand side of Eq.~\eqref{eq:v_renormalization} is related to the normalized 1$s$ exciton wave function at zero electron-hole separation (evaluated in the absence of coupling to light), since in momentum space this wave function is given by
\begin{equation}
  \Phi^{1s}_{\k} = \sqrt{\frac{2 \pi \eb }{\mu}} \frac{1}{\eb  +
    \bar{\epsilon}_{\k}}\; .
\label{eq:1s_ex}    
\end{equation}

The matter-light coupling strength $g$ can be renormalized by
considering the single-polariton problem at zero doping, and matching
the eigenvalues of the microscopic problem to the experimental
observables. Experiments typically fit the lower (LP) and upper (UP) polaritons to a coupled oscillator model (describing a tightly bound exciton and a photon):
\begin{subequations}
  \begin{align}
    \mathcal{H}_{2o} &= \begin{pmatrix} -\eb+\delta & \Omega/2 \\ \Omega/2
      & -\eb \end{pmatrix}\\
    E_{LP,UP}  &=-\eb+ \Frac{\delta \mp \sqrt{\delta^2 + \Omega^2}}{2}\;
    .
  \end{align}
\end{subequations}
Here, $\delta$ is the photon-exciton detuning and $\Omega$ the Rabi splitting. 
As such, the procedure we follow is to write the finite (renormalized) Rabi splitting $\Omega$ in terms of the bare coupling $g$ and the relative 1$s$ exciton wave function at zero electron-hole separation, which describes the amplitude for electron and hole to overlap (see~\cite{LiPRL2021} for details):
\begin{equation}
    \Omega=\frac{2g}{\area}\sum_{\k}^{\Lambda}\Phi^{1s}_{\k}=\Frac{2g}{v}
    \sqrt{\Frac{2 \pi \eb }{\mu}} \; .
\label{eq:renorm_g}    
\end{equation}
Because $1/v$ diverges logarithmically with the cutoff, $g \sim
1/\ln\Lambda \to 0$ when $\Lambda \to \infty$, in such a way that the physically meaningful
parameter $\Omega$ is finite. 

We finally turn to the photon-exciton detuning $\delta$.
Here, Ref.~\cite{LiPRB2021} found that to match the coupled-oscillator model, there is a shift from the bare detuning $\nu_{\0} + \eb$ associated with our Hamiltonian.  Specifically, one has that
\begin{equation}
    \delta=\nu_{\0} + \eb -\Frac{\Omega^2}{8\eb }\; .
\label{eq:ph_shift}    
\end{equation}

In summary, we take the finite (renormalized) energy scales in our problem to be the exciton binding energy $\eb$, and
the zero doping Rabi splitting $\Omega$. Further, the other relevant parameters are  the photon-exciton detuning $\delta$, the Fermi energy $\ef$, and the mass ratio $m_2/m_1$.

\section{Trion state}
\label{sec:spin-pwave-t}
In this section, we summarize relevant results about the $p$-wave trion, formed from indistinguishable majority particles.

\subsection{Limit of large mass ratio}
\label{sec:limit_mass_ratio}
The Hamiltonian in Eq.~\eqref{eq:Hamiltonian} is rotationally symmetric, and consequently the trion states have definite angular momentum. In the following we consider both the case of distinguishable and indistinguishable fermions, and identify the differences between these cases. To clearly illustrate the role played by exchange symmetry in determining the angular momentum, it is instructive to first consider the limit of large majority over minority mass ratio, $m_1/m_2\gg1$, when we can use the Born-Oppenheimer approximation~\cite{Born_Oppenheimer_1927,Landau:QM}.
This consists of assuming that the light particle at position $\mathbf{r}$ adiabatically adjusts its wave function to the positions of the two heavy particles at $\pm \mathbf{R}/2$, such that the total wave function takes the form
\begin{align}
\Psi(\Rvec,\rvec)=\phi(\Rvec)\psi_\Rvec(\rvec)\; .
\end{align}
The wave function of the light particle $\psi_\Rvec(\rvec)$ is obtained by solving the \sch equation for fixed $\Rvec$.
This can be shown~\cite{Ngampruetikorn_EPL2013} to have a solution in terms of the modified Bessel function of the second kind, $K_0(\kappa r)$.
Since the solution must be a parity eigenstate under $\Rvec\to-\Rvec$, there are two possibilities for $\psi_\Rvec(\rvec)$~\cite{Ngampruetikorn_EPL2013}
\begin{multline}
\psi_{\Rvec,\pm}(\rvec)= \mathcal{N}_\pm(R) \big[K_0\left(\kappa_\pm(R)\left|\rvec+\Rvec/2\right|\right)
\\
\pm K_0\left(\kappa_\pm(R)\left|\rvec-\Rvec/2\right|\right)\big]\; ,
\label{eq:psiBO}
\end{multline}
where $\mathcal{N}_\pm(R)$ is an overall normalization, and the 
momentum scale $\kappa_\pm(R)$
associated with the motion of the light particle is obtained by solving the equation $\ln(\kappa_\pm a_B)=\pm K_0(\kappa_\pm R)$ with $a_B\equiv 1/\sqrt{2\mu\eb}$~\footnote{This
  condition is obtained by applying the Bethe-Peierls boundary
  condition when the light particle approaches one of the heavy particles: $\lim_{\tilde r\to0}[\tilde r(\psi)'_{\tilde
        r}/\psi]=1/\ln(\tilde
    re^\gamma/2a_{2D})$ with $\mathbf{\tilde
      r}\equiv\rvec\pm\Rvec/2$~\cite{Ngampruetikorn_EPL2013}, and
    $\gamma\simeq0.577$ is the Euler gamma.}.  

\begin{figure}
\includegraphics[width=0.485\textwidth]{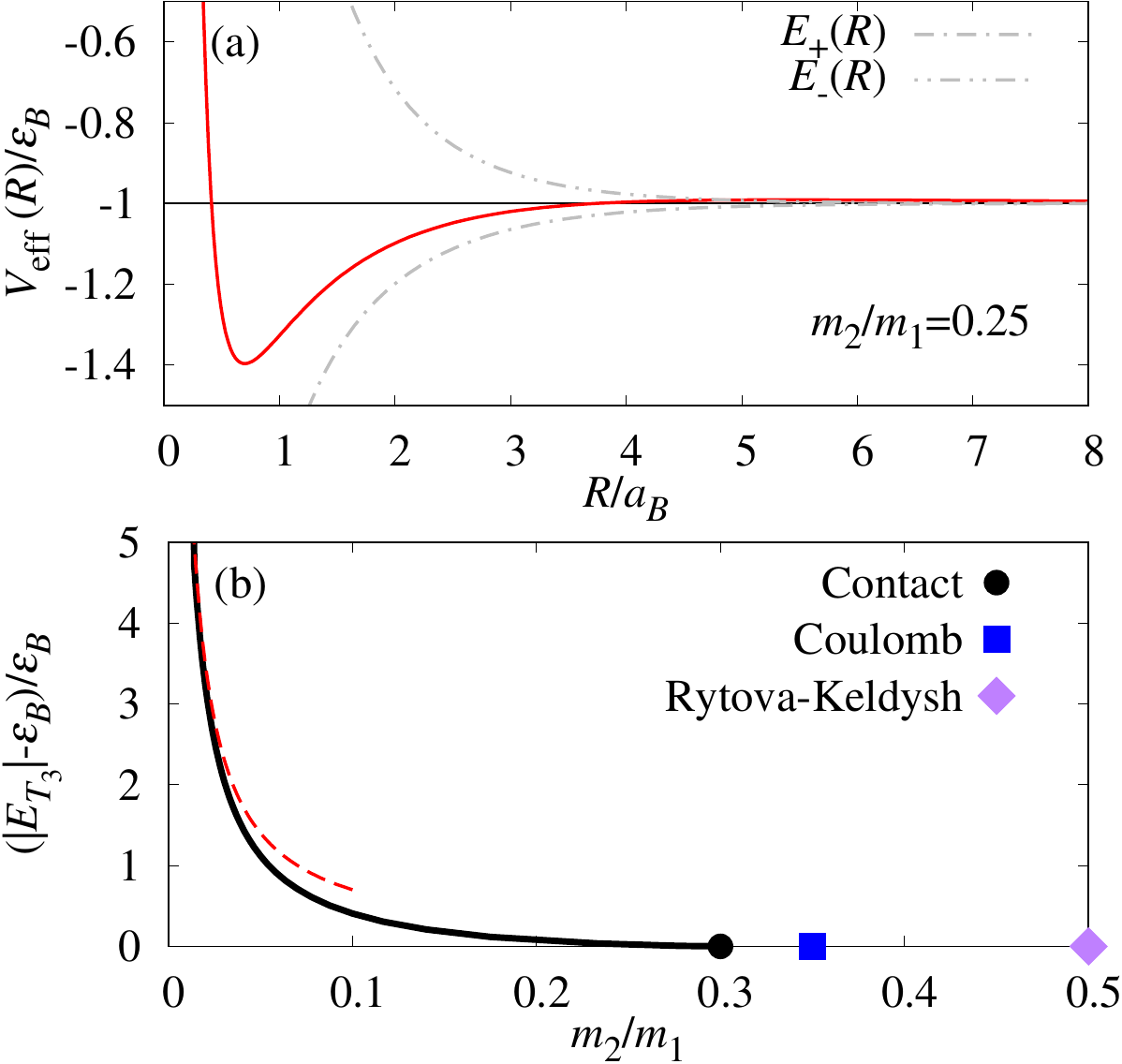}
\caption{Panel(a): Effective potentials $V_{\rm eff}(R)=E_{\pm}(R)+\ell^2/m_1R^2$ for the majority heavy particle motion in the trion state in the Born-Oppenheimer approximation for mass ratio $m_2/m_1 = 0.25$. The solid (red) line is $V_{\rm eff}(R)=E_{+}(R)+1/m_1R^2$ for the $p$-wave $\ell=\pm 1$ trion, while gray dot-dashed and dot-dot-dashed lines are $E_{+}(R)$ and $E_{-}(R)$, respectively.  Panel (b): Lowest-energy $p$-wave trion binding energy (solid black line), using a contact electronic interaction, as a function of the
minority over majority mass ratio $m_2/m_1$. The trion is bound for $m_2/m_1 \lesssim 0.3$ (circle black symbol) and thus is an $X^+$ state. At small mass ratio, the binding energy diverges according to Eq.~\eqref{eq:Hydrg_spectr} (dashed red line). 
Squared (blue) and rhombus (purple) symbols indicate the critical mass ratios obtained for bare Coulomb~\cite{Sergeev_PSS2001,Sergeev_Nanot2001} and Rytova-Keldysh  effective interactions~\cite{Courtade_PRB2017}.}
\label{fig:T3}
\end{figure}

Having solved for the motion of the light particle at a fixed separation of the heavy particles, one considers the motion of the heavy particles in the presence of the effective potential $E_\pm(R)=-\kappa_\pm^2(R)/2\mu$~\footnote{Here we use the reduced mass rather than the light particle mass~\cite{Turgut_Springer_2016}, and hence the energy surface corresponds to the energy of the relative motion of the light particle relative to one of the heavy particles. 
}
mediated by the light particle. Here, we should note that only $E_+(R)$ corresponds to a potential energy surface below the exciton at $-\eb$ (see Fig.~\ref{fig:T3}(a)), and hence only this wave function can lead to trion formation.  

In the case of distinguishable heavy particles, there are no restrictions on the overall symmetry under exchange of these particles, and so the symmetry of $\psi_{\Rvec,+}(\rvec)$ does not impose any restrictions on the symmetry of $\phi(\Rvec)$.  As such, the ground state trion is the lowest energy solution, which is  in the $s$-wave angular momentum channel.  By contrast, if the two heavy particles are identical fermions, as considered in this paper, the overall wave function must be antisymmetric under $\Rvec\to-\Rvec$. Since the attractive potential corresponds to the function $\psi_{\Rvec,+}(\rvec)$ which is symmetric under exchange, and since the total wave function $\Psi$ is antisymmetric, $\phi(\Rvec)$ must then be antisymmetric under exchange. This in turn implies that $\phi(\Rvec)$ has odd angular momentum, and therefore the ground state trion forms in the $p$-wave channel. 

The effective potentials $V_{\text{eff}}(R)=E_{\pm}(R)+\ell^2/m_1R^2$ including the centrifugal barrier for angular momentum $\ell$ are illustrated in Fig.~\ref{fig:T3}(a): We see that the bare mediated potential $E_{+}(R)$ (corresponding to the  $s$-wave case $\ell=0$) is purely attractive, while the $p$-wave potential $E_{+}(R)+1/m_1R^2$ has an attractive well when $R\sim a_{B}$, within which the trion forms, and a centrifugal barrier at small $R$. 
We also find that the attractive well and bound state only exist when the mass ratio $m_2/m_1$ is sufficiently small. The form of the potential and in particular the short-range repulsion provided by the centrifugal barrier means that the critical mass ratio for trion binding is relatively insensitive to the precise form of the interaction potential between heavy particles. 

In the limit of a large mass ratio, $m_2/m_1 \to 0$, the trion energy $E_{T_3}$ in the Born-Oppenheimer approximation takes the known form~\cite{Ngampruetikorn_EPL2013}:
\begin{equation}
    \Frac{|E_{T_3}|-\eb }{\eb } \simeq \Frac{m_1}{m_2}\Frac{2 e^{-2\gamma}}{9}\; ,
\label{eq:Hydrg_spectr}    
\end{equation}
where $\gamma \simeq 0.577$ is the Euler constant. We will now compare this with an exact calculation within our model~\eqref{eq:Hamiltonian}.

%

\subsection{Trion binding energy}
\label{sec:tri_bind_en}
Having explained why the trion ground state is $p$-wave, we next discuss the range of mass ratios for which this state is bound~\cite{Pricoupenko_PRA2010,Parish_PRA2013}. 
We thus consider a trion state in vacuum with zero center of mass momentum, which is described by the following 
state
\begin{equation}
  |T_3 \rangle= \frac{1}{\sqrt{2}\area} \sum_{\k_1 \ne \k_2}
  \gamma_{\k_1 \k_2}^{} \hat{c}^{\dag}_{-\k_1 -\k_2,2}
  \hat{c}^{\dag}_{\k_1,1} \hat{c}^{\dag}_{\k_2,1} |0\rangle \; ,
\label{eq:T3_state}
\end{equation}
where we must obey $\gamma_{\k_2 \k_1}^{} = -\gamma_{\k_1 \k_2}^{}$.
We then test whether this trion state is bound by calculating its energy and comparing with the exciton energy.
The trion
energy can be found by minimizing  $\langle T_3| (\hat{H} -E)
|T_3\rangle$ with respect to the complex 
wave function $\gamma_{\k_1 \k_2}^{*}$ to obtain the
following eigenvalue equation
\begin{equation}
  E \gamma_{\k_1\k_2}^{} = \mathcal{E}_{\k_1\k_2} \gamma_{\k_1\k_2}^{} -
  \frac{v}{\area} \sum_{\k'} \left( \gamma_{\k'\k_2}^{} +
  \gamma_{\k_1\k'}^{} \right) \; ,
\label{eq:T3_eigenvalue}
\end{equation}
where $\mathcal{E}_{\k_1\k_2} = \epsilon_{\k_1,1}+\epsilon_{\k_2,1}+\epsilon_{\k_1+\k_2,2}$.
As noted above, the ground state must have overall angular momentum $\ell=\pm1$ (i.e., it is $p$-wave), so we may consider the following ansatz:
\begin{equation}
  \gamma_{\k_1\k_2}^{} =
  e^{i\theta_1}\gamma_{k_1 k_2 (\theta_2-\theta_1)}^{} \; .
\label{eq:p-wave_ansatz}  
\end{equation}
In this case, in agreement with Refs.~\cite{Pricoupenko_PRA2010,Parish_PRA2013}, we find that the $p$-wave trion binds for 
a mass ratio $m_2/m_1 \lesssim 0.3$ (see Fig.~\ref{fig:T3}(b)). We see that our calculated binding energy closely matches that obtained within the Born-Oppenheimer approximation in the limit of a large mass ratio, Eq.~\eqref{eq:Hydrg_spectr}, see the dashed (red) line in Fig.~\ref{fig:T3}(b).

In typical semiconductors, the hole effective mass is larger than the electron mass.  As such, the critical mass ratio obtained above implies that a $p$-wave trion bound state can exist only if the majority particles are holes, meaning that it is an 
$X^+$ trion.  By contrast, for distinguishable particles, the $s$-wave bound state  exists for all mass ratios~\cite{Pricoupenko_PRA2010}, and both $X^+$ and $X^-$ trions are possible.

As noted in Sec.~\ref{sec:modelH}, we approximate the interaction between charges as a contact interaction, which can be considered as assuming charges are strongly screened.  This overestimates the $p$-wave trion binding energy when $m_2/m_1 \to 0$. In fact, the contact interaction causes the binding energy to diverge according to Eq.~\eqref{eq:Hydrg_spectr}, while an interaction that decays at large momentum transfer (like a screened Coulomb interaction) would instead result in a finite binding energy~\cite{Sergeev_Nanot2001,Sergeev_PSS2001,Courtade_PRB2017}. 
Nevertheless, the contact interaction approximation correctly describes the existence of a critical mass ratio.  Furthermore, the critical ratio found for contact interactions,
$m_2/m_1 \lesssim  0.3$ agrees well with that found for bare Coulomb interactions~\cite{Sergeev_PSS2001,Sergeev_Nanot2001},  $m_2/m_1 \lesssim 0.35$, as well as  for Rytova-Keldysh effective interactions describing TMD monolayers~\cite{Courtade_PRB2017},  $m_2/m_1 \lesssim 0.5.$ These critical mass ratios are marked by symbols in Fig.~\ref{fig:T3}(b).

\subsection{Coupling to light}
\label{sec:couple_light}
The dipole matrix element for the transition between a single isolated trion and a single carrier vanishes---i.e., the isolated trion does not couple to light.
This can be easily shown by considering the matrix element of the matter-light interaction term $\hat{H}_{ehC}$~\eqref{eq:H_ehC} between the trion 
state in Eq.~\eqref{eq:T3_state} and a cavity photon plus a majority particle state at zero momentum $|C+1\rangle = \hat{a}^{\dag}_{\0} \hat{c}^{\dag}_{\0,1} |
  0\rangle$~\cite{Glazov_JCP2020}:
\begin{equation}
  \langle T_3|\hat{H}_{ehC} |C+1\rangle = \frac{\sqrt{2}g}{\sqrt{\area}}\frac1\area
  \sum_{\k}\gamma_{\k\0}^* \; .
\label{eq:T3+ph_state}
\end{equation}
This term is in general vanishingly small due to the prefactor of order $1/\sqrt{\area}$~\cite{Combescot_SSC2003}; such a suppression is present for both $s$- and $p$-wave trions.
Additionally, in the $p$-wave case we have $\frac1\area\sum_{\k}\gamma_{\k\0}^* = 0$ due to
Eq.~\eqref{eq:p-wave_ansatz},
i.e., the $p$-wave transition is further forbidden by symmetry.

Now let us discuss how these results lead to a non-zero oscillator strength in experiments carried out at small but finite doping. In the $s$-wave case, the oscillator strength of the trion branch is proportional to the square of the matrix element in Eq.~\eqref{eq:T3+ph_state} multiplied by the number of particles within the area $\area$, i.e., the oscillator strength scales as the majority particle density $\sim n_1$. This estimate is smaller than the exciton oscillator strength by a factor proportional to $n_1$, in agreement with results based on the trion~\cite{Esser_pssb_2001,Glazov_JCP2020,zhumagulov2021microscopic} and the polaron~\cite{Combescot_SSC2003,Sidler_NP2016,Efimkin_PRB2017} pictures. 

In order to estimate the finite-density trion oscillator strength in the $p$-wave case, we need to consider additionally the correction to the sum appearing in Eq.~\eqref{eq:T3+ph_state} since this is identically zero at $k_F=0$. For $k_F\ll \sqrt{2\mu |E_{T_3}|}$, we therefore instead calculate the matrix element between a final-state trion at a typical center of mass momentum $\mathbf{k}_F\equiv k_F \hat{\mathbf n}$ (where $\hat{\mathbf{n}}$ is a unit vector in an arbitrary direction), and an initial state with a photon at normal incidence and a carrier at momentum $\mathbf{k}_F$. To leading order, the matrix element becomes $(\sqrt{2}g/\area^{3/2})\sum_{\k}\gamma^*_{\k \k_F}$. It is straightforward to show that, in the $p$-wave case, this sum scales linearly with $k_F\propto \sqrt{n_1}$. To see this, we can rewrite Eq.~\eqref{eq:T3_eigenvalue} in terms of $\eta_\k=\frac v\area\sum_{\k'}\gamma_{\k'\k}$~\footnote{In practice, having a Fermi sea restricts the momentum $\k_1$ appearing in the sums to be above the Fermi surface. However, due to the $p$-wave symmetry, such corrections are higher order in $k_F$ and may be neglected here.}:
\begin{equation}
    \left(\frac1v+\frac1\area\sum_{\k_1}\frac1{E-\mathcal{E}_{\k_1\k_2}}\right)\eta_{\k_2}=\frac1\area\sum_{\k_1}\frac{\eta_{\k_1}}{E-\mathcal{E}_{\k_1\k_2}}.
\label{eq:implicit_trion-eq}
\end{equation}
Like $\gamma$, $\eta$ satisfies the $p$-wave symmetry $\eta_{\k}^{} =
e^{i\theta}\eta_{k} $. Therefore, the right hand side is identically zero at $\k_2=\0$, and thus $\eta_\0 =0$
and the matrix element in Eq.~\eqref{eq:T3+ph_state} vanishes. At finite doping, instead, we want to estimate $\eta_{\k_F}$ at small $k_F$. Expanding the kinetic energy $\mathcal{E}_{\k_1\k_F}$ for small $\k_F$ to linear order, and using the $p$-wave condition, we then find the first non-zero term scales as $k_F$, multiplied by a $k_F$-independent integral. Taking the square amplitude of the matrix element and multiplying by the number of majority particles within the area $\area$, we thus find that in the $p$-wave case the trion oscillator strength $\sim n_1^2$.
This estimate agrees with the numerical results obtained within the polaron picture, as analyzed below in Sec.~\ref{sec:spec_vs_dope}.

\section{Exciton polaron polariton state}
\label{sec:M4}
In this section, we present a wave function ansatz  describing exciton-polaron (polariton) states with indistinguishable carriers---also denoted as the ICP case---as sketched in Fig.~\ref{fig:schematic}(a). We first present the ansatz, which we use to describe both ground and excited states, and then discuss how we may efficiently calculate the absorption spectrum within this ansatz.

\subsection{Wave function ansatz and eigenvalue equations}
We consider the following state, inspired by the Chevy ansatz~\cite{Chevy2006}, which describes a superposition of a bare photon and an exciton with a dressing cloud of electron-hole excitations of the Fermi sea, all with zero center of mass momentum:
\begin{multline}
  |\widetilde{M}_4\rangle= \Bigg( \alpha \hat{a}^{\dag}_{\0} + \sum_{\k_1}
  \frac{\varphi_{\k_1}^{}}{\sqrt{\area}} \hat{c}^{\dag}_{-\k_1,2} \hat{c}^{\dag}_{\k_1,1}\\
  +
  \sum_{\k_1, \k_2, \q} \frac{\varphi_{\k_1\k_2\q}^{}}{\sqrt{2} \area^{3/2}}
  \hat{c}^{\dag}_{\q-\k_1-\k_2,2} \hat{c}^{\dag}_{\k_1,1}
  \hat{c}^{\dag}_{\k_2,1} \hat{c}_{\q,1}^{} \Bigg) |FS\rangle \; ,
\label{eq:M4_state}
\end{multline}
normalized so that $1 = \langle\widetilde{M}_4|\widetilde{M}_4\rangle = |\alpha|^2 + \frac{1}{\area} \sum_{\k_1} |\varphi_{\k_1}^{}|^2 + \frac{1}{\area^3} \sum_{\k_1, \k_2, \q} |\varphi_{\k_1 \k_2 \q}^{}|^2$.
In Eq.~\eqref{eq:M4_state}, $|FS\rangle = \prod_{\q} \hat{c}_{\q,1}^{\dag} |0\rangle$ describes the Fermi sea  of majority particles, and 
we use the convention that momenta labeled $\k_i$  represent states above the Fermi sea ($k_i>\kF$), while momenta labeled $\q$ refer to states below ($q<\kF$).  We denote the state in Eq.~\eqref{eq:M4_state} by $\widetilde{M}_4$~\cite{Parish_PRA2013} to indicate it is a molecular (i.e., excitonic) state with up to four-body correlations.
The first two terms in  $|\widetilde{M}_4\rangle$  are, respectively, a photon with amplitude $\alpha$, and an electon-hole pair (undressed exciton) with wave function $\varphi_{\k}^{}$ in terms of the relative electron-hole momentum $\k$.  The final term describes a scattered exciton and a single intra-band particle-hole excitation of the majority particle Fermi sea. The associated four-body wave function $\varphi_{\k_1\k_2\q}^{}$ can be viewed as a trion-like (three-particle) complex plus a hole of the Fermi sea. Indeed, this term reduces to the trion wave function~\eqref{eq:T3_state} in the limit of vanishing doping. Thus, we will refer to this term as the ``trion-hole'' state for brevity, although it should be understood that the ``trion'' in this complex is not necessarily a well-defined three-particle bound state. 
Because majority particles are indistinguishable, the trion-hole wave function $\varphi_{\k_1\k_2\q}^{}$ must be antisymmetric under the exchange $\k_1$ and $\k_2$, i.e., $\varphi_{\k_1\k_2\q}^{} = -\varphi_{\k_2\k_1\q}^{}$, which is satisfied by all our numerical results in the following.
%

In our variational state, Eq.~\eqref{eq:M4_state}, we consider only the states where the photon is at zero momentum, as these are experimentally accessible by a probe at normal incidence. Furthermore, we do not include the contribution of the particle-hole dressed photon state,
\begin{equation}
    \sum_{\k\q} \Frac{\alpha_{\k\q}^{}}{\area} \hat{a}^{\dag}_{\q-\k} \hat{c}^{\dag}_{\k,1} \hat{c}_{\q, 1} | FS\rangle \; .
\label{eq:dressed_photon}    
\end{equation}
If present, this term would lead to a broadening of the photon, because a photon at $\Q=\0$ could scatter to a different momentum $\q-\k$, which is typically non-zero because $q<\kF$ and $k>\kF$. 
However, due to the photon mass $m_C$ being approximately five orders of magnitude smaller than the bare electron mass ($m_C \simeq 10^{-5} m_0$), finite-momentum photons have energies far off-resonance with both trion and exciton energies.  As such, their contribution is negligible~\cite{Levinsen_PRL2019}. 

The wave function $|\widetilde{M}_4\rangle$ can be considered as a variational ansatz for the lowest energy 
state. It can however also describe a truncated basis within which to consider excited states~\cite{Parish2016}, and thus the manifold of states that are relevant for optical absorption.  
To see this, we minimize the expectation value $\langle \widetilde{M}_4| (\hat{H} -E)
|\widetilde{M}_4\rangle$ with respect to the complex variational parameters $\alpha^*$, $\varphi_{\k}^{*}$, and $\varphi_{\k_1 \k_2 \q}^{*}$ to obtain the
following eigenvalue equations:
\begin{widetext}
\begin{subequations}
\begin{align}
\label{eq:M4_eigenvalue_photon}
E \alpha &= \nu_{\0}\alpha -
  \frac{g}{\area}\sum_{\k_1'}\varphi_{\k_1'}^{}\\
  \label{eq:M4_eigenvalue_exciton}
  E \varphi_{\k_1}^{} &= \bar{\epsilon}_{\k_1}
  \varphi_{\k_1}^{}-\frac{v}{\area}\sum_{\k_1'} \varphi_{\k_1'}^{} -
  g\alpha
  -\frac{\sqrt{2} v}{\area^2}\sum_{\k_2'\q'} \varphi_{\k_1\k_2'\q'}^{}
  \\
\label{eq:M4_eigenvalue_2}
  E \varphi_{\k_1\k_2\q}^{} &= \mathcal{E}_{\k_1\k_2\q}
  \varphi_{\k_1\k_2\q}^{} + \frac{v}{\area}\sum_{\q'}
  \varphi_{\k_1\k_2\q'}^{} - \frac{v}{\area}\sum_{\k_1'}
  \varphi_{\k_1'\k_2\q}^{} - \frac{v}{\area}\sum_{\k_2'} \varphi_{\k_1\k_2'\q}^{}-
  \Frac{v}{\sqrt{2}}\left(\varphi_{\k_1}^{} - \varphi_{\k_2}^{} \right) \; ,
\end{align}
\label{eq:M4_eigenvalue}
\end{subequations}
\end{widetext}
where, as in Eq.~\eqref{eq:v_renormalization}, $\bar{\epsilon}_{\k_1}=\epsilon_{\k_1,1}+\epsilon_{\k_1,2} =
\k_1^2/2\mu$, while $\mathcal{E}_{\k_1\k_2\q} = \epsilon_{\k_1,1} + \epsilon_{\k_2,1} -
\epsilon_{\q,1} + \epsilon_{\q-\k_1-\k_2,2}$.
By solving the coupled linear equations~\eqref{eq:M4_eigenvalue} we gain direct access to the energies of both the ground and excited states, as well as the corresponding wave functions. This includes information about the photon, exciton, and dressed exciton amplitudes.

Equations~\eqref{eq:M4_eigenvalue} show that the photon mode couples only to the bare exciton part of the state.  However, the bare exciton and trion-hole terms are not system eigenstates in the presence of doping; they couple  via the electron-hole interaction.  As such---when the $p$-wave trion state is bound, i.e., for $m_2/m_1 \lesssim 0.3$---the eigenstates are hybridized to form attractive and repulsive polaron resonances, as well as an incoherent continuum of many-body states.
Because of this hybridization, the photon couples to all of these eigenstates, leading to a transfer of oscillator
strength from the repulsive branch---which, at low doping, is exciton-like---to the attractive branch---which, at low doping, is  trion-like. 
  
Even though for indistinguishable particles the symmetry of the (three-body) trion state is $p$-wave~\eqref{eq:p-wave_ansatz}, the lowest energy wave functions of both the (two-body) exciton and (four-body) trion-hole contributions have an overall $s$-wave symmetry:
\begin{subequations}
\label{eq:M4_Ansatz}
\begin{align}
  \varphi_{\k_1}^{} &=  \varphi_{k_1}^{}\\
  \varphi_{\k_1\k_2\q}^{} &= \varphi_{k_1 k_2 q (\theta_1-\theta_q)
    (\theta_2-\theta_q)}^{}\; .
\label{eq:M4_Ansatz_2}    
\end{align}
\end{subequations}
Indeed, within our model Hamiltonian~\eqref{eq:Hamiltonian}, all other angular momentum states are completely uncoupled to light, and hence we will be making this $s$-wave ansatz  in the following.
Clearly,
in the case of the trion-hole wave function $\varphi_{\k_1\k_2\q}^{}$, Eq.~\eqref{eq:M4_Ansatz_2} implies that we can equivalently choose as reference angle either the angle of the Fermi sea hole $\theta_q$, as in Eq.~\eqref{eq:M4_Ansatz_2}, or the angle of any of the two majority species particles, e.g., $\varphi_{\k_1\k_2\q}^{} = \varphi_{k_1 k_2 q (\theta_2-\theta_1)
    (\theta_q-\theta_1)}^{}$.
Note that, at low doping, the trion sub-space within the trion-hole complex of those states corresponding to the attractive branch still has an angular momentum $\ell=\pm1$, while the hole component has $\ell=\mp1$, as shown later in Fig.~\ref{fig:ang_momentum} and in Appendix~\ref{app:T3h_ang-momentum}. Yet, as we discuss below, the orbital character of both attractive and repulsive branches
evolves with doping.

In solving the system of equations in~\eqref{eq:M4_eigenvalue}, we want to consider the limit where the UV cut-off $\Lambda\to\infty$, and replace the bare parameters $v,g,\nu_0$ with the renormalized parameters, as discussed in Sec.~\ref{sec:renormalization}. 
Our results will then be independent of microscopic physics, and can be expressed in terms of the exciton binding energy $\eb$, the photon-exciton detuning $\delta$, and the Rabi splitting $\Omega$. In addition to these, the other relevant parameters are the Fermi energy $\ef$ and the mass ratio $m_2/m_1$.
By considering the limit
$\Lambda \to \infty$, $v \to 0$, we may also simplify the form of Eq.~\eqref{eq:M4_eigenvalue_2}. Since the large-$k$ behavior of both exciton and trion-hole wave functions is $\varphi_{\k_{1,2}}^{} \sim  \varphi_{\k_1,\k_2,\q}^{}  \sim 1/k_{1,2}^2$, we can neglect the term $\frac{v}{\area}\sum_{\q'}
  \varphi_{\k_1\k_2\q'}^{}$ in Eq.~\eqref{eq:M4_eigenvalue_2} when $\Lambda \to \infty$.
  
We numerically solve the eigenvalue equations~\eqref{eq:M4_eigenvalue} by discretizing the momenta on a grid---see Appendix~\ref{app:numerical} for details. 
For efficiency, we use a grid in polar coordinates that exploits the symmetry of the system.
Note that this approach is only possible because of our use of contact interactions.  If we had used full Coulomb interactions, this inevitably also requires intraspecies interactions (which vanish in the contact case).  Such intraspecies interactions lead to the appearance of terms in the eigenvalue equations that involve differences of momenta, and, as such, do not lie on the original momentum grid.

Finally, we note that while Eq.~\eqref{eq:M4_eigenvalue} is written allowing for a strong light-matter coupling, $g$, it can also be considered in the limit $g\to 0$. 
This therefore allows us to explore two distinct regimes: In the strong coupling regime, the 2D semiconductor is embedded in a microcavity and the coupling to light explicitly modifies the excitonic states resulting in the formation of polaron-polaritons. Conversely, in the weak coupling regime, the 2D semiconductor is probed by light in the absence of a cavity, and the probe light does not change the form of the spectrum. Technically, in our formulation, the latter case corresponds to removing the photonic part of the variational state in Eq.~\eqref{eq:M4_state} (corresponding to removing Eq.~\eqref{eq:M4_eigenvalue_photon}) and taking $g=0$ in Eq.~\eqref{eq:M4_eigenvalue_exciton}, and thus this procedure will be implicit in the following whenever we discuss results obtained in the weak coupling limit.

\subsection{Spectral functions}
\label{sec:GreenF}
A natural probe of trion states is optical absorption which 
can be calculated, in both strong- and weak-coupling regimes, starting from  the photon and ``exciton'' Green's functions respectively. 
In the time-domain, these are defined as
\begin{align}
    G_{C,X} (t) &= \langle \Psi_0^{C,X}| e^{-i \hat{H} 
    t} |\Psi_0^{C,X}\rangle\; .
\end{align}
Here, $|\Psi_0^{C,X}\rangle$ denotes the initial state within the space spanned by our ansatz.  The choice of this state varies depending on which Green's function we seek to calculate.  For the photon Green's function $(C)$ we consider an initial state with a single photon
\begin{equation}
    |\Psi_0^C\rangle = \hat{a}^{\dag}_{\0} |FS\rangle\; .
\end{equation}
The ``exciton'' Green's function ($X$) is instead chosen to describe the response of the material to optical excitation, and thus the initial state we use is that of an electron-hole pair at the same spatial position---note that this state is \emph{not} an exciton.  We will nonetheless refer to this as the exciton Green's function in the following, since this name is commonly used in the literature. We thus write  
\begin{equation}
|\Psi_0^{X}\rangle = 
    \displaystyle 
  \Frac{\mathcal{N}}{\sqrt{\area}} \sum_{\k} \hat{c}^{\dag}_{-\k_,2} \hat{c}^{\dag}_{\k_,1} |FS\rangle \; ,
\end{equation}
where the normalization $\mathcal{N}=(\frac{1}{\area} \sum_\k)^{-1/2}$ is chosen so that $\langle \Psi_0^X|\Psi_0^X\rangle=1$.

Both Green's functions can be written in the frequency domain in terms of the complete set of eigenstates of
Eqs.~\eqref{eq:M4_eigenvalue}, described by eigenvalues $E_n$, and photonic $\alpha^{(n)}$ and excitonic $\varphi_{\k}^{(n)}$ components of the eigenvectors:
\begin{subequations}
\label{eq:both_Greens-f}
\begin{align}
\label{eq:C_Greens-f}  
  G_{C} (\omega) &= \sum_n \Frac{|\alpha^{(n)}|^2}{\omega -
    E_n + i\epsilon}\\
    G_{X,\Lambda} (\omega) &= \sum_n \Frac{\left|\frac{\mathcal{N}}{\area} \sum_{\k} \varphi_{\k}^{(n)} \right|^2}{\omega -
    E_n + i\epsilon}  \; ,
\label{eq:X_Greens-f}
\end{align}
\end{subequations}
where $\epsilon$ denotes a Lorentzian linewidth which we add by hand.
We note that the exciton Green's function $G_{X,\Lambda}$ defined in Eq.~\eqref{eq:X_Greens-f} depends on the UV cut-off $\Lambda$ and needs to be renormalized in order to obtain a physical quantity which is cut-off independent.
This is because as  $\Lambda \to \infty$, $\mathcal{N} \sim \Lambda^{-1} \to 0$ while $\frac{1}{\area} \sum_{\k} \varphi_{\k}^{(n)} \sim \log \Lambda \to \infty$.
One can show that a cut-off independent form can be obtained by considering the following rescaling
\begin{align}
  G_{X} (\omega) &=\left( \Frac{2g}{\Omega}\right)^2 \Frac{G_{X,\Lambda} (\omega)}{\mathcal{N}^2} \nonumber\\
  &=\sum_n \Frac{\left|\frac1\area\sum_\k\varphi^{(n)}_\k\Big/\frac1\area\sum_\k\Phi^{\rm 1s}_\k \right|^2}{\omega -
    E_n + i\epsilon} \; ,
\label{eq:X_Greens-f_renorm}
\end{align}
where the microscopic matter-light coupling constant $g$ and the Rabi splitting $\Omega$ are related by Eq.~\eqref{eq:renorm_g}, and $\Phi^{\rm 1s}_\k$ denotes the wavefunction of the 1$s$ exciton~\eqref{eq:1s_ex}. With this definition, it is easy to see that $G_{X} (\omega)$ is cut-off independent when $\Lambda \to \infty$.

Because the photon mass is orders of magnitude  smaller than that of the exciton, we can neglect the electron-hole photon dressing term in Eq.~\eqref{eq:dressed_photon}. 
This simplifies the problem considerably, in particular we find that the strong coupling photon Green's function in Eq.~\eqref{eq:C_Greens-f} is related to the exciton Green's function in the weak light-matter coupling regime, which we denote $G^{(0)}_{\text{X}}(\omega)$,
via~\footnote{In the limit of tightly bound excitons, the same observation about the form of the Green's function for small photon mass has been derived in Ref.~\cite{Levinsen_PRL2019}.}:
\begin{equation}
    G_{C} (\omega) = \frac{1}{\omega -\delta + \eb - (\Omega/2)^2 G_{\text{X}}^{(0)} (\omega) +i\epsilon}\; .
\label{eq:strong-and-weak-c}    
\end{equation}
Note however that $G^{(0)}_{\text{X}}(\omega)$ is not the free exciton Green's function,
because of the electron-hole dressing by the Fermi sea.

If we try to evaluate the exciton and photon Green's functions by first finding the complete set of eigenvalues and eigenvectors of Eqs.~\eqref{eq:M4_eigenvalue}, this places significant constraints on the number of degrees of freedom we may consider, and thus on the precision of the calculation.
%
We thus also employ a recursive method, originally developed by Haydock and collaborators~\cite{Haydock_JPC1972}, which allows one to consider a larger number of basis states, and thus reach a higher numerical precision. 
Here, one seeks to transform the eigenvalue problem~\eqref{eq:M4_eigenvalue} into a tridiagonal form, with the top left element corresponding to the expectation value on the initial state $|\Psi^{C,X}_0\rangle$.  Truncating this recursive scheme at some order provides a basis of those states which are most relevant in terms of their contribution to the associated Green's function.
The Green's function can then be conveniently evaluated by continued-fraction.
For our problem we can simultaneously evaluate both the photon Green's function $G_{C} (\omega)$, as well as the renormalized exciton Green function in the weak coupling limit to light 
$G^{(0)}_{\text{X}}(\omega)$, which are related by Eq.~\eqref{eq:strong-and-weak-c}.  

We are interested in the optical absorption both in weak and strong coupling.  As such, we define two spectral functions as follows:
\begin{subequations}
\label{eq:spectral-f}
\begin{align}
\label{eq:spectral-f_X}
    W_{X} (\omega) &= -\Frac{1}{\pi} \Im  G_{X}^{(0)} (\omega) \\
    W_{C} (\omega) &= -\Frac{1}{\pi} \Im G_{C} (\omega)\; .
\end{align}
\end{subequations}
These functions have different meanings.
The exciton spectral function $W_X(\omega)$ corresponds to the absorption by the semiconductor (TMD monolayer or quantum well), in the absence of any optical microcavity.  The relation between the photon spectral function $W_C(\omega)$ and optical absorption is more subtle.  This spectral function does correspond to the absorption of light by an optical microcavity containing the semiconductor, but only in the limit where the cavity linewidth is much smaller than that of the excitons~\footnote{When the cavity linewidth is comparable or greater than that of the semiconductor, optical absorption requires considering an input-output approach~\cite{Ciuti2006}.}.

\begin{figure}
\includegraphics[width=0.5\textwidth]{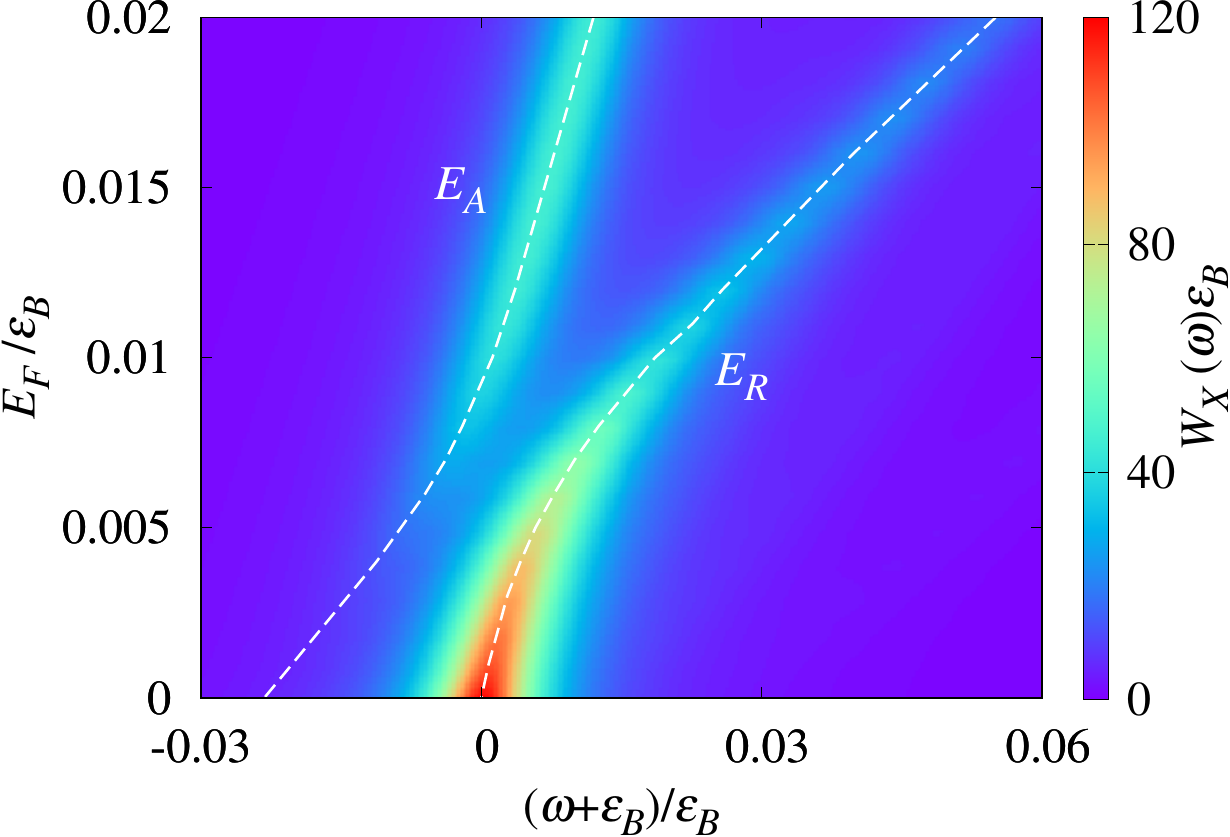}
\caption{Colormap of the exciton spectral function $W_{X} (\omega)$ for the ICP case, in the weak coupling regime 
as a function of the frequency $\omega$ and the majority particle Fermi energy $\ef$. 
The spectrum  displays Lorentzian resonances (see Fig.~\ref{fig:M4_sec}), corresponding to attractive ($\omega=E_A$) and  repulsive ($\omega=E_R$) polaron branches (indicated by dashed white lines). At $\ef \to 0$, these approach the trion ($E_A \to E_{T_3}$) and exciton ($E_R \to -\eb $) energies, respectively.
At the mass ratio used, $m_2/m_1=0.25$, the $p$-wave trion binding energy is $(|E_{T_3}| - \eb )/\eb  \simeq 0.022$ (see Fig.~\ref{fig:T3}(b)). 
The linewidth is $\epsilon=5\times 10^{-3}\eb $.}
\label{fig:M4_map}
\end{figure}
\begin{figure}
\includegraphics[width=0.5\textwidth]{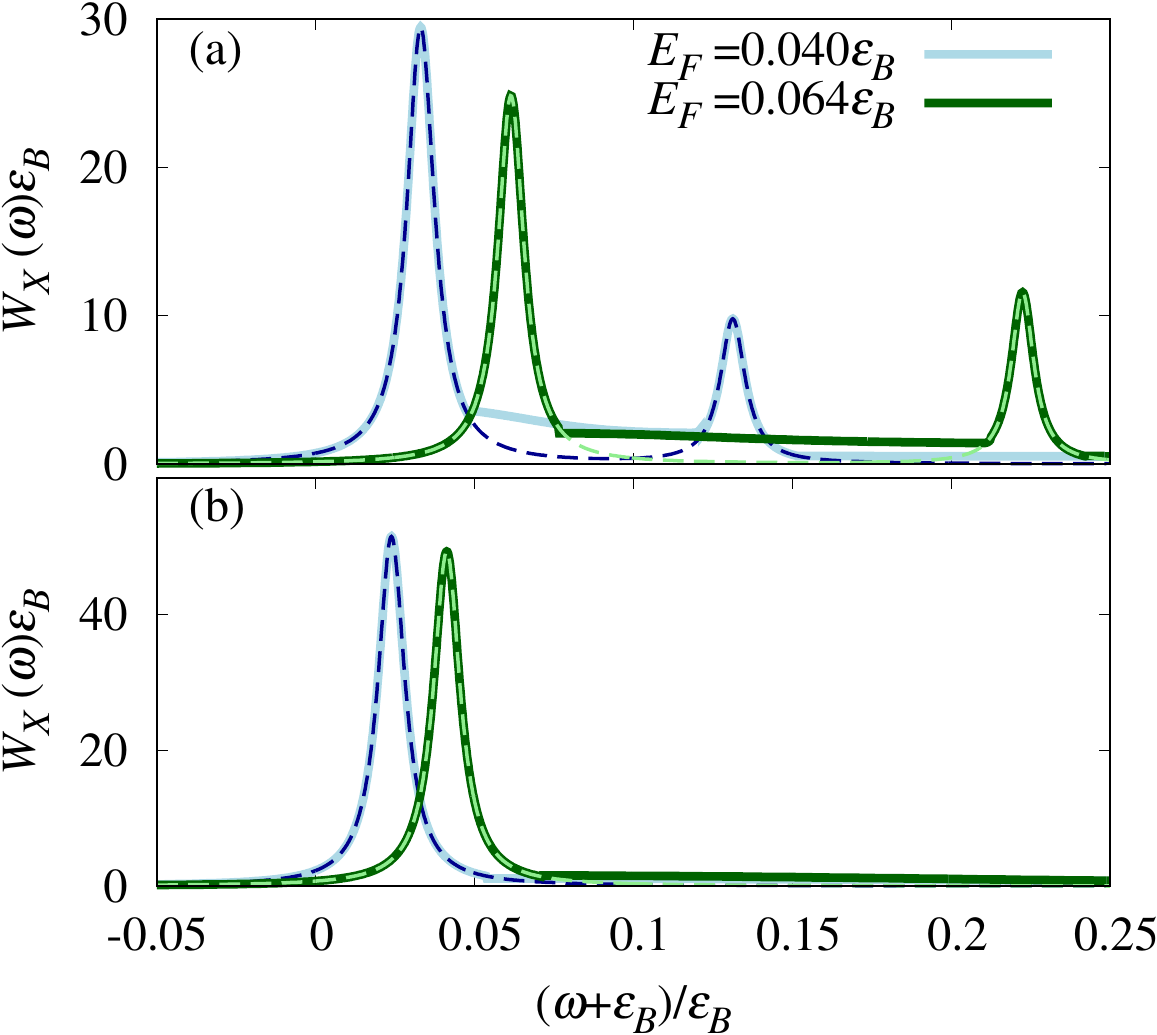}
\caption{Cross section of the convolved exciton spectral function  $W_X(\omega)$ for the ICP case in the weak coupling regime, 
plotted for two different values of $\ef$, and two values of mass ratio: (a) $m_2/m_1=0.25$, (b) $m_2/m_1=1$.  
 Dashed lines are a Lorentzian fit, Eq.~\eqref{eq:Lorentz_fit},  to the spectra. 
In panel (a) the $p$-wave trion is bound and the spectrum displays both attractive and repulsive branches, while in  panel (b) the $p$-wave trion is unbound and the spectrum has a single peak.
The linewidth is $\epsilon=5\times 10^{-3}\eb $. 
To produce a smooth spectrum, the Green's function is convolved---details of this procedure can be found in Appendix~\ref{app:numerical}.
} 
\label{fig:M4_sec}
\end{figure}
%
\subsection{Weak coupling}
%
\subsubsection{Spectral function for indistinguishable carriers}
We first present our results in the weak coupling regime. This corresponds to the absence of a cavity, and therefore we 
consider the exciton spectral function $W_{X} (\omega)$.
This function shows distinct behavior depending on whether the $p$-wave trion state is bound ($m_2/m_1 \lesssim 0.3$) or not ($m_2/m_1 > 0.3$)---see Sec.~\ref{sec:tri_bind_en}. 
When the $p$-wave trion state is bound, the spectral function is characterized by two peaks, as shown in  Fig.~\ref{fig:M4_map} and Fig.~\ref{fig:M4_sec}(a). We identify these as the attractive branch at $\omega = E_A$ and the repulsive branch at $\omega = E_R$ (the location of these peaks are indicated by dashed white lines in Fig.~\ref{fig:M4_map}).  In the limit of zero doping $\ef \to 0$, the attractive mode continuously connects with the $p$-wave trion state ($E_A \to E_{T_3}$), while the repulsive mode tends toward the exciton state ($E_A \to -\eb $). 
As anticipated, in this limit, the trion has vanishing spectral weight, and the spectral function has a single peak at the exciton energy with spectral weight (i.e., integrated area) equal to one.
Upon increasing the doping, we see that there is a transfer of spectral weight from the repulsive (exciton) to the attractive (trion) branch.

\begin{figure*}
\includegraphics[width=1.0\textwidth]{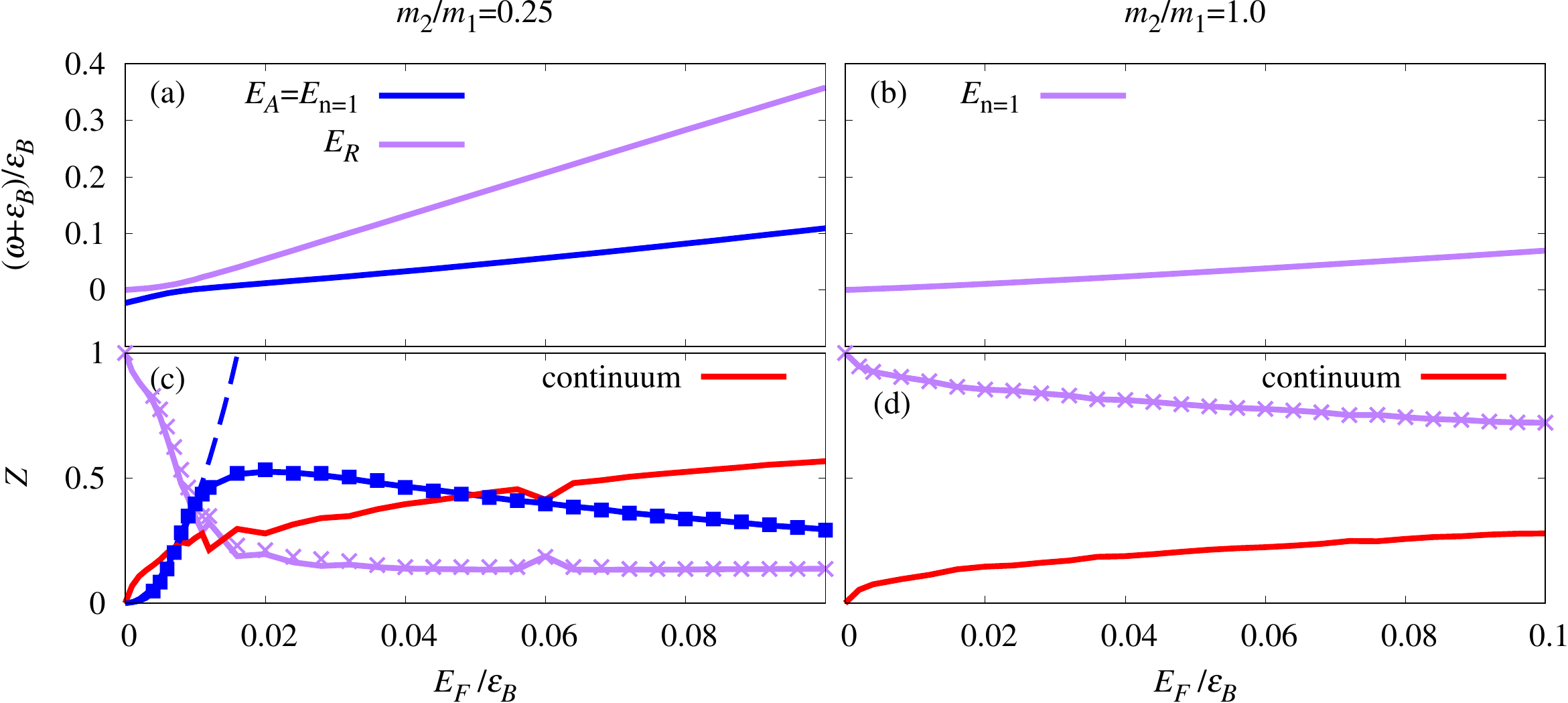}
\caption{Doping dependence of the branch energies (a,b) and spectral weights (c,d) of the exciton spectral function $W_X(\omega)$ in the weak coupling regime 
for the ICP case. For mass ratio  $m_2/m_1=0.25$ (left panels), the spectrum has two peaks: the attractive branch  at $\omega = E_A= E_{n=1}$ (solid blue line), which coincides with the lowest eigenvalue of Eqs.~\eqref{eq:M4_eigenvalue}, and the repulsive branch at $\omega = E_R$ branch (solid purple line). 
For $m_2/m_1=1$ (right panels), the spectrum is characterized by a single peak at $\omega = E_{n=1}$ (solid purple line).
The solid lines correspond to the spectral weights found by integrating the peak areas for the R,A branches, and using $Z_{\text{continuum}}=1-Z_R-Z_A$ for the continuum weight.  
In panel (c), the dashed blue line shows a fit to the attractive branch weight with $Z_A \sim (\ef/\eb)^{1.92}$.
The symbols (blue squares and purple crosses) show the spectral weight extracted from results in the strong coupling limit obtained for $\Omega = 0.2\eb$,  fitting the  polariton spectra to the  three coupled oscillator model in Eq.~\eqref{eq:3_coupled}, thus extracting the Rabi splittings $\Omega_{A,R}$, and then using  $Z_{A,R}= (\Omega_{A,R}/\Omega)^2$.
}
\label{fig:M4_properties}
\end{figure*}

The 
energy of the  attractive branch peak coincides with the lowest eigenvalue of Eqs.~\eqref{eq:M4_eigenvalue}, $E_A = E_{n=1}$, and we find that there is a strong suppression of spectral weight immediately above this value. As such, the attractive branch at finite doping retains a Lorentzian shape, which mirrors the DCP case~(see, e.g., Ref.~\cite{Goulko2016} for a detailed discussion in the three-dimensional case).
Conversely, the repulsive branch does not coincide with a single eigenvalue, because of the presence of a 
continuum in between the attractive and repulsive modes. Despite this, we find that the repulsive branch also has a Lorentzian shape with a constant width $\epsilon$ for all values of doping $\ef$. This can be clearly observed in Fig.~\ref{fig:M4_sec}(a), where the spectral function $W_X (\omega)$ is plotted for two values of the Fermi energy $\ef$.  As such, we conclude that, for dressing by indistinguishable carriers, the shape of the repulsive branch is not affected by the 
continuum. We explain this result below in  Sec.~\ref{sec:spec_vs_dope}, by showing that at large enough doping the repulsive branch and the 
continuum have distinct symmetries and, as such, do not hybridize. Note that the ICP and DCP cases are very different in this regard, as discussed in Sec.~\ref{sec:comparison}.

\subsubsection{Evolution of spectral weight with doping}
\label{sec:spec_vs_dope}
We next explore how the attractive and repulsive peak positions and weights evolve with doping.  
To do this, we fit the weak coupling exciton spectral function  $W_X (\omega)$ with two Lorentzians centered at $E_{A,R}$ and with quasiparticle weights $Z_{A,R}$:
\begin{equation}
  W_X (\omega) \simeq -\frac1\pi {\rm Im}\left[\Frac{Z_A}{\omega - E_A + i\epsilon} + \Frac{Z_R}{\omega - E_R + i\epsilon}\right]\; .
\label{eq:Lorentz_fit}
\end{equation}
The weights $Z_{A,R}$ correspond to the areas underneath the peaks. 
Examples of these fits are shown Fig.~\ref{fig:M4_sec}(a) (dashed lines), illustrating that these peaks fit this Lorentzian form extremely well.

In addition to the Lorentzian peaks, the spectral function also includes the continuum of many-body 
states. 
We denote the weight of this continuum as $Z_{\text{continuum}}$, and we estimate its value from a sum rule on the exciton Green's function. 
In fact, numerically we find that 
the 
exciton Green's function satisfies:
\begin{equation}
    -\Frac{1}{\pi}\int_{-\infty}^{E_N^{\0}} d\omega \Im G^{(0)}_{X} (\omega) = 1\; ,
\label{eq:sum-rule}
\end{equation}
where the integral is up to the energy $E_N^{\0} = k_F^2/2\mu$ of an unbound majority-minority pair on top of a Fermi sea with zero centre of mass motion $\Q=\0$.
Because of this, one can write that:
\begin{equation}
 Z_A + Z_R + Z_{\text{continuum}} = 1  \; .
\end{equation}

Figure~\ref{fig:M4_properties} shows
the doping dependence of the energies of attractive and repulsive branches $\omega = E_{A,R}$, and of the weights $Z_{A,R}$ and $Z_{\text{continuum}}$, plotted for the same two mass ratios as shown in Fig.~\ref{fig:M4_sec}.
We observe that both attractive and repulsive branches are blue-shifted when $\ef$ increases.  The blue-shift of the upper (repulsive) branch can naturally be understood from repulsion between the levels. The blue-shift of the attractive branch can be understood as arising from the Pauli exclusion experienced by  the optically generated majority particle.
As doping first increases,  the spectral weight of the repulsive branch is transferred to both the continuum and the attractive branch, however initially the attractive branch weight grows more slowly. 
Eventually, for $\ef \gtrsim 0.02 \epsilon_B$, both attractive and repulsive branches transfer their weights to the continuum. Note that the weights  $Z_{A,R}$ coincide with the oscillator strengths of attractive and repulsive modes, and thus, in the strong coupling regime, determine the Rabi splittings $\Omega_{A,R}=\sqrt{Z_{A,R}}\Omega$ of the polariton modes, as discussed further in Sec.~\ref{sec:strong}.
Note also that the repulsive and continuum spectral weights have a slightly noisy behavior for specific values of $\ef$. As explained in Appendix~\ref{app:numerical}, this is a finite-size effect of the numerical calculation.

\begin{figure}
\includegraphics[width=0.5\textwidth]{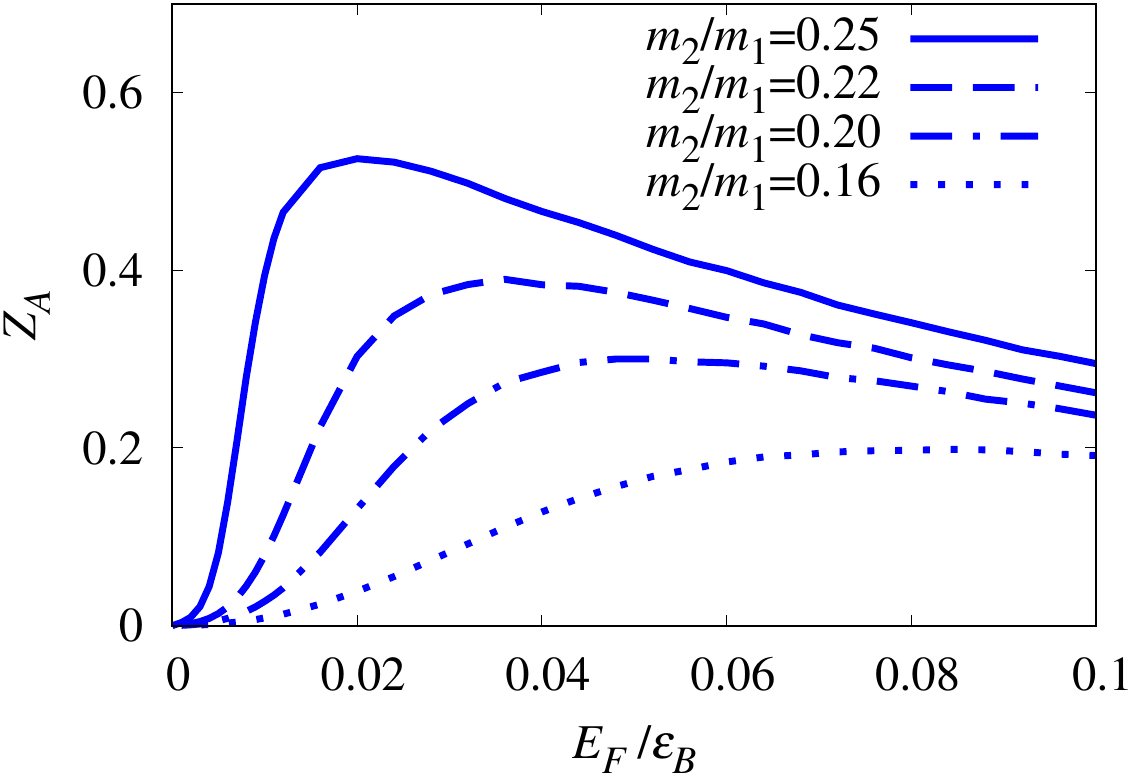}
\caption{Spectral weight of the attractive branch $Z_A$ as a function of the Fermi energy $\ef$ for different values of the minority over majority mass ratios $m_2/m_1$ for the ICP case.}
\label{fig:A_branch_mass-ratio}
\end{figure}

We find that the growth of the attractive branch spectral weight with doping is consistent with quadratic, with a fit to our numerical results giving $Z_A \sim (\ef/\eb)^{1.92}$ (dashed blue line in Fig.~\ref{fig:M4_properties}(c)). This is quite different from the behavior known for the DCP case, where the growth is linear~\cite{Glazov_JCP2020,zhumagulov2021microscopic}.  
This different power-law dependence of $Z_A$ on $\ef$ can be understood directly from the difference of $s$-wave and $p$-wave symmetry of the  trion state belonging to the trion-hole complex.  In both cases, one factor of $\ef$ dependence arises to account for the relative probability of creating a majority electron-hole pair, as discussed in Ref.~\cite{Glazov_JCP2020}.   In the $p$-wave case, an extra factor arises since, as discussed in Sec.~\ref{sec:couple_light}, the matrix element to create a trion  from an electron at 
$\k=\0$ vanishes by symmetry.  As such, the amplitude for the transition to a trion-hole state depends not only on the density of carriers, but on a momentum-weighted density, giving a higher power of $\ef$.

The dependence of $Z_A$ on $\ef$ varies with the mass ratio.
This is shown in Fig.~\ref{fig:A_branch_mass-ratio}, where we plot the spectral weight of the attractive branch $Z_A$ vs $\ef$ for a variety of mass ratios $m_2/m_1<0.3$.
The transfer of spectral weight is reduced at larger mass imbalance, when the $p$-wave trion is more strongly bound.

For $m_2/m_1 > 0.3$, the $p$-wave trion is unbound and, as noted above, the spectral function displays a single peak. This single branch continuously connects, at zero doping, with the exciton mode $E_{n=1} \to -\eb $, as illustrated in Fig.~\ref{fig:M4_properties}(b). The weights of  the $E_{n=1}$ state and the continuum are plotted in Fig.~\ref{fig:M4_properties}(d), showing  a gradual weight transfer from the single branch to the continuum when $\ef$ increases. This transfer to the continuum is slower than when the trion $p$-wave state is bound.
We have also analyzed the case where the trion is unbound at zero doping and becomes bound  by increasing $\ef$~\cite{Parish_PRA2013}. Surprisingly, we find that the spectral function displays two branches exclusively when the trion state is already bound at $\ef=0$ and otherwise is characterized by the repulsive branch only.

\begin{figure}
\includegraphics[width=0.5\textwidth]{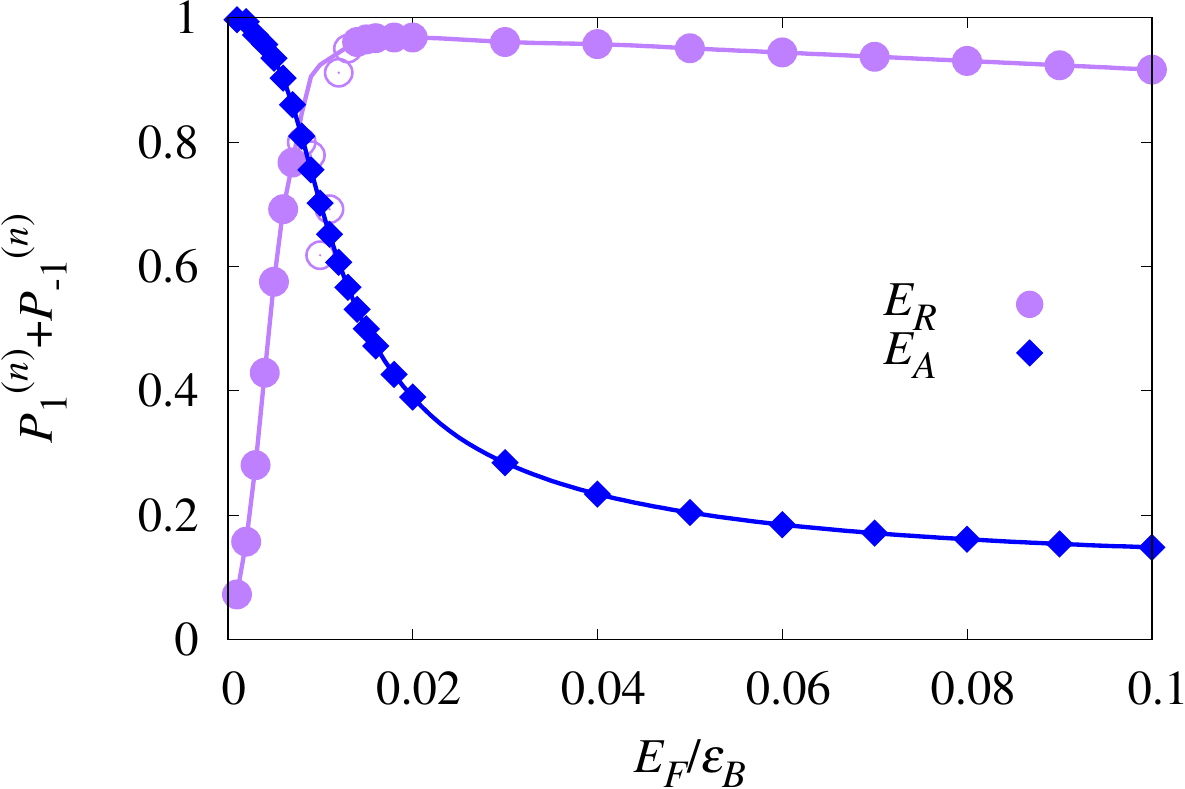}
\caption{Doping dependence of the orbital character of the repulsive and attractive branches of the ICP case at $m_2/m_1=0.25$. We show the orbital character by plotting $P_{\ell=1}^{(n)}+P_{\ell=-1}^{(n)}=2P_{\ell=1}^{(n)}$, the probability to find the majority-species hole in the trion-hole complex in angular momentum channel $\ell=\pm 1$, i.e., having $p$-wave character---see Eq.~\eqref{eq:hole-ang-mom_prob}  for the formal definition. Symbols denote the values found from exact diagonalization.  For the repulsive branch (purple) we use empty symbols to indicate values where a resonance with one of the continuum states occurs; the location and form of this resonance is strongly dependent on finite-size effects as discussed in Appendix~\ref{app:T3h_ang-momentum}.  Solid lines are a guide to the eye, excluding these points.
We plot this character for the eigenstate $n$ closest to the attractive $E_A$ and repulsive $E_R$  branches.}
\label{fig:ang_momentum}
\end{figure}

As noted previously, in our ansatz, the four-body complex described by the wave function $\varphi_{\k_1\k_2\q}^{}$ in the ansatz~\eqref{eq:M4_state} always has an overall $s$-wave symmetry, $\ell=0$. The three-particle (trion) and Fermi sea hole subspaces within the complex can however have any orbital character consistent with this, i.e., the overall state can be a superposition of states where the trion and hole have opposite angular momenta $\ell_{\text{trion}}=-\ell_{\text{hole}}$; in practice we find that components with $\ell_{\text{hole}}=0,\pm 1$ dominate the state. In order to evaluate the hole (and, consequently, the trion) angular momentum in the trion-hole complex, we consider
the probability $P_{\ell}^{(n)}$ for the hole in a given eigenstate $n$ to have an angular momentum $\ell$---for the precise definition, see Appendix~\ref{app:T3h_ang-momentum}.
In Fig.~\ref{fig:ang_momentum}, we plot the doping dependence of $P_{\ell=1}^{(n)}+P_{\ell=-1}^{(n)}=2P_{\ell=1}^{(n)}$ for the eigenvalue $E_n$ closest to the attractive $E_A$ and repulsive $E_R$  branches.  
When this quantity is zero,  the hole (and trion) in the trion-hole term is $s$-wave, while when it is one the hole and trion are $p$-wave. 

We observe that, as expected, at zero doping the attractive branch hole (and thus trion) has a $p$-wave symmetry, consistent with earlier arguments about the ground state of the trion for indistinguishable carriers.
In the same limit,  the repulsive branch hole (and thus trion) has an $s$-wave symmetry.
At very low doping, when the attractive branch spectral weight $Z_A$ has a quadratic dependence on $\ef$, the attractive branch is primarily $p$-wave and  the repulsive branch $s$-wave. However, at larger doping this switches to a regime  where the attractive branch becomes instead $s$-wave and the repulsive branch $p$-wave.
Note that, as discussed in  Appendix~\ref{app:T3h_ang-momentum}, the optically active continuum always retains the $s$-wave symmetry. When the continuum has the same or larger spectral weight of the repulsive branch, as seen at $\ef \gtrsim 0.02 \eb$, the repulsive branch is fully $p$-wave. The different symmetry of the repulsive branch versus the continuum explains why the repulsive branch retains its Lorentzian shape with constant width $\epsilon$, and does not hybridze with the continuum. This behavior for the ICP case is very different than that seen for the DCP case, as we discuss next.

\begin{figure}
\includegraphics[width=0.5\textwidth]{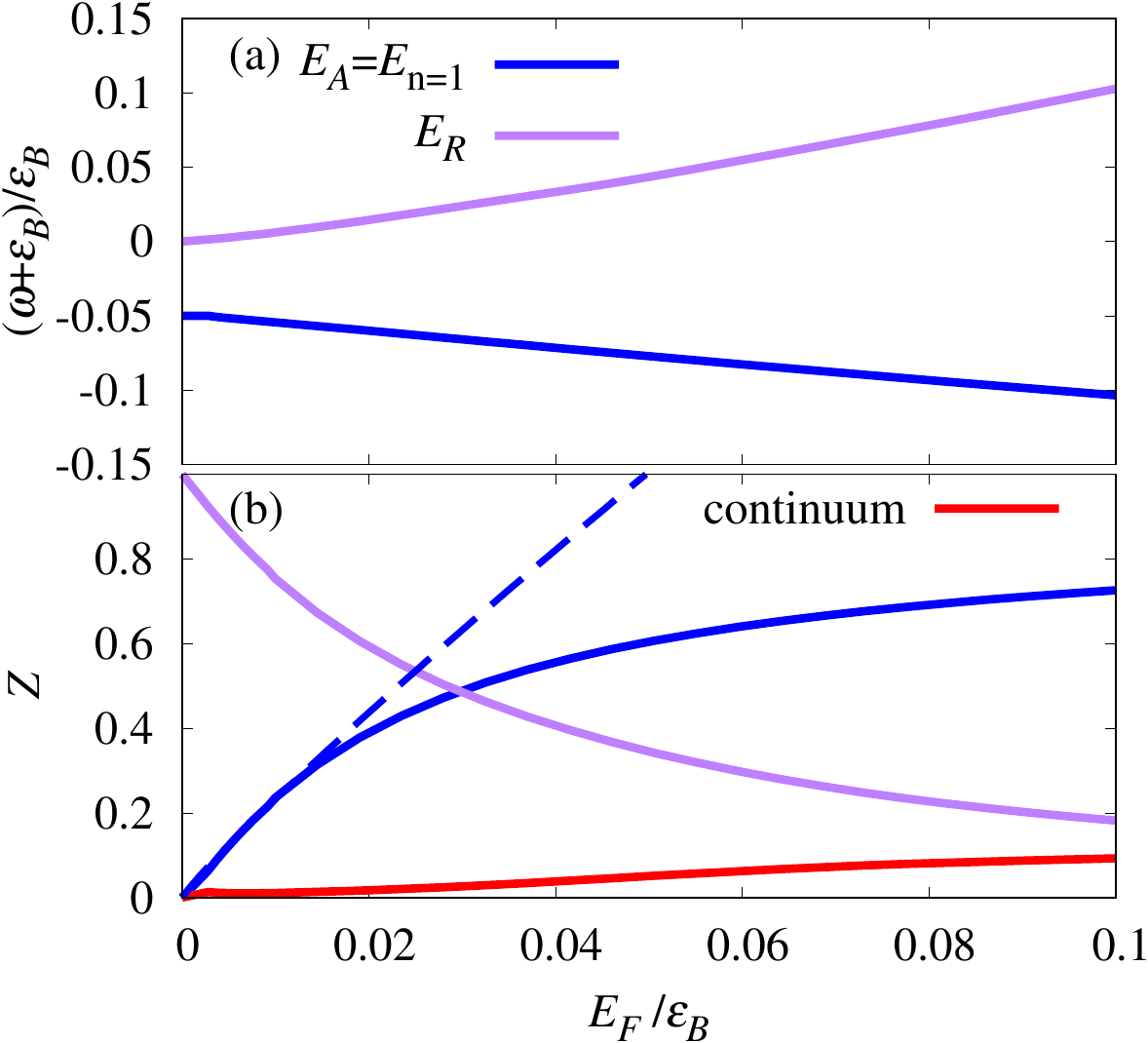}
\caption{Doping dependence of energies and spectral weights obtained for the DCP case where the Fermi sea is distinguishable from the carriers in the exciton, so that the zero doping trion is $s$-wave~\cite{Efimkin_PRB2017,Sidler_NP2016,Rana_PRL2021,Efimkin_PR2021}.
Details of this model and the parameters used  are given in Appendix~\ref{app:distinguish}.
Panel (a): Doping dependence of the attractive $E_A$ and repulsive $E_R$ peak positions. Panel (b): Spectral weights of attractive and repulsive branches, as well as the 
continuum (solid red line). A fitting of the attractive branch weight with $Z_A \sim (\ef/\eb)^{0.93}$ is shown as a dashed blue line.}
\label{fig:spin-swave_polaron_properties}
\end{figure}
%
\subsubsection{Comparison to the DCP case}
\label{sec:comparison}
It is instructive to contrast the indistinguishable carrier case described above, where the trion has a $p$-wave symmetry, with  the case of dressing by distinguishable carriers, where the trion state is $s$-wave.  The DCP case has previously been studied in depth~\cite{Sidler_NP2016,Efimkin_PRB2017,Rana_PRL2021,Efimkin_PR2021}. For completeness, details of this model are given in Appendix~\ref{app:distinguish}. 

In the DCP case, the $s$-wave trion is always bound, i.e., $|E_{T_3}|-\eb>0$, for any mass ratio~\cite{Sergeev_PSS2001,Sergeev_Nanot2001,Courtade_PRB2017}. 
%
Therefore, the spectrum is also characterized by attractive and repulsive branches which continuously connect to the $s$-wave trion and exciton states, respectively. As in the ICP case, the attractive branch is well separated from the continuum 
and thus has a Lorentzian shape. However, the repulsive branch is in this case hybridized with the 
continuum and so its shape is not Lorentzian. Instead it has an asymmetric shape, and a linewidth that grows with $\ef$~\cite{Sidler_NP2016,Efimkin_PR2021}. 

Figure~\ref{fig:spin-swave_polaron_properties} shows the doping dependence of the energy and quasiparticle weights in the DCP case, which can be directly compared with the corresponding results in the ICP case shown in Fig.~\ref{fig:M4_properties}.
%
From the energy peaks (panel (a)) one sees that the attractive branch here red-shifts with doping; this difference is because, for distinguishable carriers, there is no effect of Pauli blocking on the exciton.
From the spectral weights (panel (b)), one sees that
the transfer to the continuum here is negligible.
Further,  at small densities, the attractive branch spectral weight has a linear dependence on density, with a fit to our numerical results giving $Z_A \sim (\ef/\eb)^{0.93}$ (dashed [blue] line), as already predicted by Refs.~\cite{Glazov_JCP2020,zhumagulov2021microscopic}.

\begin{figure}
\includegraphics[width=0.5\textwidth]{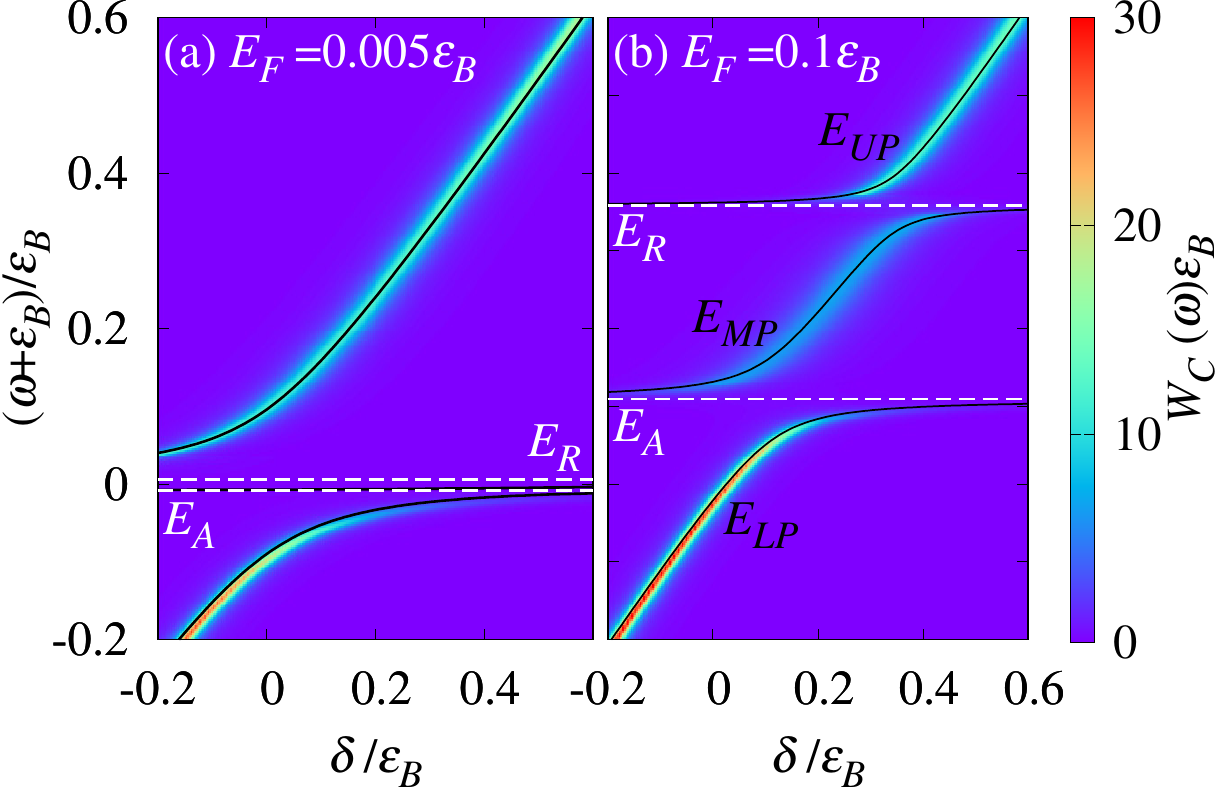}
\caption{Photon spectral function $W_C (\omega)$ for the ICP case in the strong light-matter coupling regime, as a function of the photon-exciton detuning $\delta$ and the frequency $\omega$, for two different values of $\ef$. Attractive ($\omega = E_A$) and repulsive ($\omega = E_R$) branches evaluated in the weak coupling regime are plotted as dashed (white) lines. The photon spectra reveals 
lower (LP), middle (MP), and upper (UP) polariton branches. The solid (black) lines are the results of the fitting with the three-oscillator model of Eq.~\eqref{eq:3_coupled}. The mass ratio is $m_2/m_1=0.25$, the Rabi splitting $\Omega = 0.2 \eb$, and the linewidth $\epsilon=10^{-2} \eb $. }
\label{fig:polariton_map}
\end{figure}

\subsection{Strong coupling}
\label{sec:strong}
We finally discuss how the above results obtained for the ICP case affect the spectrum in the regime of strong light-matter coupling.
In Fig.~\ref{fig:polariton_map} we plot the photon spectral function $W_C (\omega)$ as a function of the photon-exciton detuning $\delta$ and the frequency $\omega$, for two different values of $\ef$, and for a mass ratio $m_2/m_1=0.25$ at which the zero doping $p$-wave trion is bound---the same conditions as Figs.~\ref{fig:M4_map}(a) and~\ref{fig:M4_properties}(a,c).
At low doping (left panel), the attractive branch has a negligible spectral weight and the attractive and repulsive branches are very close to each other, and therefore we  see only two polariton branches. At larger $\ef$ (right panel), we instead see three polariton branches, the
lower (LP), middle (MP), and upper (UP) polariton. This occurs because the oscillator strength transfer from the repulsive to the attractive branch allows for anticrossings of the photon with both 
branches.

We can associate a Rabi splitting to each exciton-polaron branch, $\Omega_{A,R}$, by fitting the three polariton branches with the eigenvalues of a three-coupled oscillator model:
\begin{equation}
  \mathcal{H}_{3o} = \begin{pmatrix} -\eb + \delta & \Omega_{A}/2 &
    \Omega_{R}/2\\ \Omega_{A}/2 & E_{A}  & 0 \\ \Omega_{R}/2 & 0 &
    E_{R}  \\
  \end{pmatrix}\; .
\label{eq:3_coupled}  
\end{equation}
We take $E_{A,R}$ and $\delta$ as fixed parameters, and then extract $\Omega_{A,R}$ as fitting parameters. The result of this fitting is shown in Fig.~\ref{fig:polariton_map} as solid lines. 
%
These Rabi splittings relate to the spectral weight via $\Omega_{A,R}=\sqrt{Z_{A,R}} \Omega$~\cite{Kohstall2012},
and a measurement of $\Omega_{A,R}$ therefore allows one to extract the corresponding quasiparticle weights that one would have in the absence of strong light-matter coupling.
The values of $Z_{A,R}$ extracted by fitting of the polariton branches are shown as symbols (squares for $Z_A$ and crosses for $Z_R$) in Fig.~\ref{fig:M4_properties}. These clearly  agree perfectly with $Z_{A,R}$ obtained from the exciton spectral function in the weak coupling regime.
We find that this procedure remains accurate for values of $\epsilon \ll \Omega\lesssim2\eb$.

\section{Conclusions and perspectives}
\label{sec:Conc}
We have studied the optical properties of a doped 2D semiconductor, where one of the two charges forming the exciton is  indistinguishable from those forming the Fermi sea induced by doping---a case we referred to  as the ICP case. 
We have calculated the optical absorption, which describes
transitions between the system ground state and states with an inter-band particle-hole pair (exciton).
To describe the effects of the Fermi sea, we employed a polaron description where the exciton is dressed by a single intra-band particle-hole excitation
of the Fermi sea.  

The polaron formalism allows us to recover, at low doping, the properties of few-body complexes, i.e., the exciton and the trion. At the same time, this formalism allows one to describe the higher density many-body regime.
From the comparison of our results with those obtained in the distinguishable (or DCP) case, we conclude that, while for the DCP case the spectral function is  always characterized by attractive and repulsive branches (because the associated $s$-wave trion is always bound), for the ICP case there are two branches only when the $p$-wave trion is bound, which requires sufficiently large majority to minority mass ratio.

Both the ICP and DCP cases show a transfer of oscillator strength from the  repulsive to the attractive branch as one increases doping.  
Such a transfer of weight is possible because, in both cases, it is not the trion state itself which must couple to light, but rather a trion-hole complex (a complex consisting of three particles and a Fermi-sea hole) that indirectly couples to light via its coupling to the exciton.
The spectral weight of the attractive branch has a different dependence on doping for the ICP and DCP cases: at low doping, in the DCP case it grows linearly with the Fermi sea density~\cite{Glazov_JCP2020,zhumagulov2021microscopic},
while in the ICP case we find that it grows quadratically
as a consequence of the 
$p$-wave nature of the trion state.
In the regime of strong light-matter coupling, 
the transfer of oscillator strength to the attractive branch furthermore leads to the appearance of 
three polariton modes resulting from the anticrossing of the photon with both the attractive and repulsive branches. We have discussed how the Rabi splittings in the strong-coupling polariton spectrum allows one to effectively measure the weak-coupling quasiparticle weights.

The attractive polaron energy recovers, at low doping, the $p$-wave ($s$-wave) trion energy for the ICP (DCP) case. In both cases, this branch 
is a sharp Lorentzian-like peak of the spectral function, with a linewidth that does not change with doping.
The repulsive polaron branch continuously connects at low doping with the exciton energy. Because attractive and repulsive branches are separated by a 
continuum, the repulsive branch never coincides precisely with a system eigenstate. However, for the ICP case, the repulsive branch is, like the attractive branch, a sharp peak with a Lorentzian shape and a doping-independent broadening. This is in stark contrast with the DCP case, where the repulsive peak is a broad feature, involving multiple eigenstates, and has an asymmetric shape and a linewidth that increases with doping~\cite{Sidler_NP2016,Efimkin_PR2021}. 
The origin of the different nature of the repulsive branch in the ICP and DCP cases comes from the orbital character of the states involved.  
We show this by calculating the angular momentum of the three-particle (trion) and Fermi-sea hole components of the polaronic state, both for the repulsive branch and for the continuum.
For the ICP case, at large enough doping, the repulsive branch and continuum states have different 
orbital characters and, thus, do not mix. For the ICP case we also observe that the orbital characters of attractive and repulsive branches swap as one increases doping, so that the Fermi-sea hole in the attractive branch has $s$-wave symmetry at high doping. 

Observing the predictions of this work requires the $p$-wave 
trion state to be bound. As noted earlier, this state is bound when the mass ratio between majority and minority particles is sufficiently large.  For smaller mass ratio, a bound trion could be achieved in a sufficiently strong out-of-plane magnetic field, in the limit where the carrier orbital motion undergoes Landau-level quantization~\cite{Sandler1992,Macdonald1992two,Dzyubenko1994,Palacios_PRB1996,whittaker_PR1997,Shields_PRB1995,Finkelstein_PRB1996}. Preliminary results~\cite{Sanvitto_PRL2002} have already indicated that, in presence of a magnetic field and by increasing doping, the exciton oscillator strength can transfer to the $p$-wave 
trion state. Extending our results  to the Landau quantized regime would be an interesting subject for future studies.

\acknowledgments We are grateful to R. Haydock and
A. H. MacDonald for useful discussions. AT and FMM acknowledge financial support from the Ministerio de Ciencia e Innovaci\'on (MICINN), projects No.~AEI/10.13039/501100011033 (2DEnLight) and
No.~MAT2017-83772-R (QLMC-2D).
FMM acknowledges financial support from the Proyecto Sinérgico CAM 2020 Y2020/TCS-6545 (NanoQuCo-CM).
JL and MMP acknowledge support from the
Australian Research Council Centre of Excellence in Future Low-Energy
Electronics Technologies (CE170100039). JL and MMP are also supported through
the Australian Research Council Future Fellowships FT160100244 and FT200100619, respectively. JK
acknowledges financial support from EPSRC program ``Hybrid
Polaritonics'' (EP/M025330/1).

\appendix

\section{Numerical evaluation of eigenvalues and spectral function}
\label{app:numerical}
Here we give details about the numerical procedure employed to determine the spectral properties of the indistinguishable carrier case. 
\subsection{Haydock iteration to calculate Green's function}
We have already commented in the main text about the numerical limitations  on obtaining the complete set of eigenvalues and eigenvectors of Eqs.~\eqref{eq:M4_eigenvalue} by exact diagonalization due to the large Hilbert space.
Furthermore, not all eigenstates are required to evaluate the spectral function.  What is thus desired is an approach to identify those states that are of most relevance to the spectral function.
For this reason, we employ the recursive method developed by Haydock and collaborators and described in Ref.~\cite{Haydock_JPC1972}.  This allows us to derive the photon and exciton Green's functions, and thus the associated spectral functions~\eqref{eq:spectral-f}, without requiring diagonalization. By starting from an appropriate initial state, this method iteratively transforms the original eigenvalue problem in Eq.~\eqref{eq:M4_eigenvalue} into a tridiagonal form, from which one can conveniently evaluate the Green's functions as a continued-fraction---see Ref.~\cite{Haydock_JPC1972} for details.

\subsection{Momentum grid and convergence}
\label{app:mom_grid}
If one were working with a finite UV cut-off, $\Lambda$, then the discrete form of the Hamiltonian describing the eigenvalue problem~\eqref{eq:M4_eigenvalue} could be obtained by considering  a grid for momenta $k \in [\kF,\Lambda]$.
However, as we wish to consider the renormalized problem, $\Lambda \to \infty$, the grid for $k$-integrals has to extend up to infinity.
We thus apply a transformation $\beta =\tan k$, with $\beta \in [\arctan(\kF), \pi/2)$. We then consider a  Gauss-Legendre quadrature in $\beta$ with $N_k$ points, in $q$ with $N_q$ points, and in $\theta$ with $N_{\theta}$ points. Note that, in this way, by sending the number of points $N_k \to \infty$, we automatically consider the  $\Lambda \to \infty$ limit.

\begin{figure}
    \includegraphics[width=0.5\textwidth]{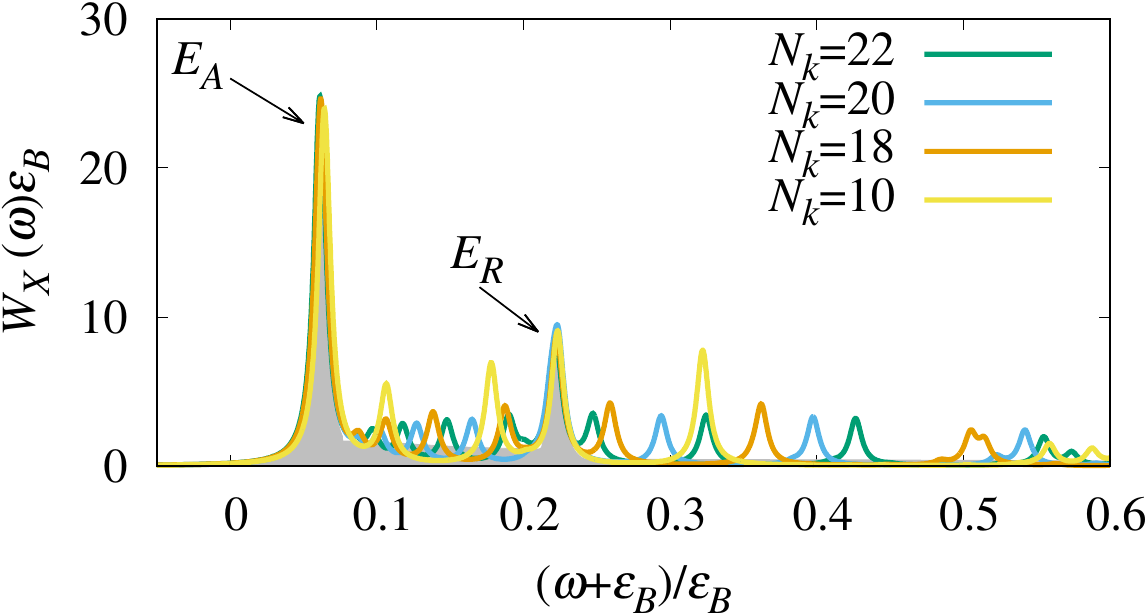}
    \caption{Exciton spectral function profiles $W_X(\omega)$ in the 
    weak coupling regime for the same system parameters as Fig.~\ref{fig:M4_sec}(a) ($\ef=0.064\eb$, $m_2/m_1=0.25$, and $\epsilon=5\times 10^{-3} \eb$), for different values of $N_k$ and for $N_q=4$ and $N_\theta=8$. The (gray) shaded area shows the convolved profile plotted in Fig.~\ref{fig:M4_sec}(a).}
    \label{fig:M4_convergence}
\end{figure}

By studying the dependence of the spectral functions on the number of points, $N_k$, $N_q$, and $N_{\theta}$, we find that the convergence with respect to $N_q$ and $N_{\theta}$ is reached easily (already for $N_q=4$, $N_{\theta}=7$), while the details of the spectra strongly depend on $N_k$, as shown in Fig.~\ref{fig:M4_convergence}. 
To be concrete, we observe that the two pronounced peaks corresponding to the attractive and repulsive polaron branches converge quickly with $N_k$.  By contrast, the set of states that eventually will form a 
continuum  continue to vary with $N_k$. This distinct behavior as a function of $N_k$ allows us to distinguish the attractive and repulsive branches from the 
continuum.
Note that we have checked that our results match between direct diagonalization of Eqs.~\eqref{eq:M4_eigenvalue} and the Haydock iteration method. The recursive method however allows us to consider a larger number of grid points (up to $N_k=22$) than the direct diagonalization method (restricted to $N_k=10$) because it only involves matrix-times-vector operations and thus requires less memory. 

Even though the recursive method allows us to reach larger values of $N_k$ than exact diagonalization, at the maximum value we can reach, $N_k=22$, the form of the 
continuum has still not converged. 
We observe in Fig.~\ref{fig:M4_convergence} that, by increasing $N_k$, the states associated to the 
continuum reduce in frequency and accumulate in the region between the attractive and repulsive branches. During this evolution, there are specific values of $N_k$ where a given continuum state becomes resonant with the repulsive branch.  Since the coupling between these modes is small, the repulsive branch does not notably shift in energy at these resonances, but it does change its spectral weight. These resonances result in the slightly ``noisy'' behavior of the spectral weight of the repulsive branch shown in Fig.~\ref{fig:M4_properties}(c).
As we are unable to predict the $N_k \to \infty$ evolution of the continuum states and are not interested in the exact shape of the continuum spectral function, we
smooth the continuum states by
applying a Gaussian convolution with a varying width,
\begin{equation}
    \Frac{1}{\sqrt{2\pi}\sigma(\omega)}\int d\omega'G_X^{(0)}(\omega')e^{-\Frac{(\omega-\omega')^2}{2\sigma(\omega)^2}}\; .
\end{equation}
Here, we choose the linewidth $\sigma(\omega)$ so that to smooth the continuum, leaving unaltered the attractive and repulsive branches. In particular, we take $\sigma(\omega)=(E_R-E_A)/2$ 
for frequencies between the repulsive and attractive branches, while $\sigma(\omega)=(E_{N}^{\0}- E_R)/2$ for frequencies above the repulsive branch.
This approach modifies the form of the continuum, but does not change its spectral weight.
Figs.~\ref{fig:M4_map},~\ref{fig:M4_sec},~\ref{fig:M4_properties},~\ref{fig:A_branch_mass-ratio}, and~\ref{fig:polariton_map} of the main text are obtained with the recursive method with $N_k=20$, $N_q=4$, and $N_\theta=8$.

\begin{figure}
    \centering
    \includegraphics[width=0.5\textwidth]{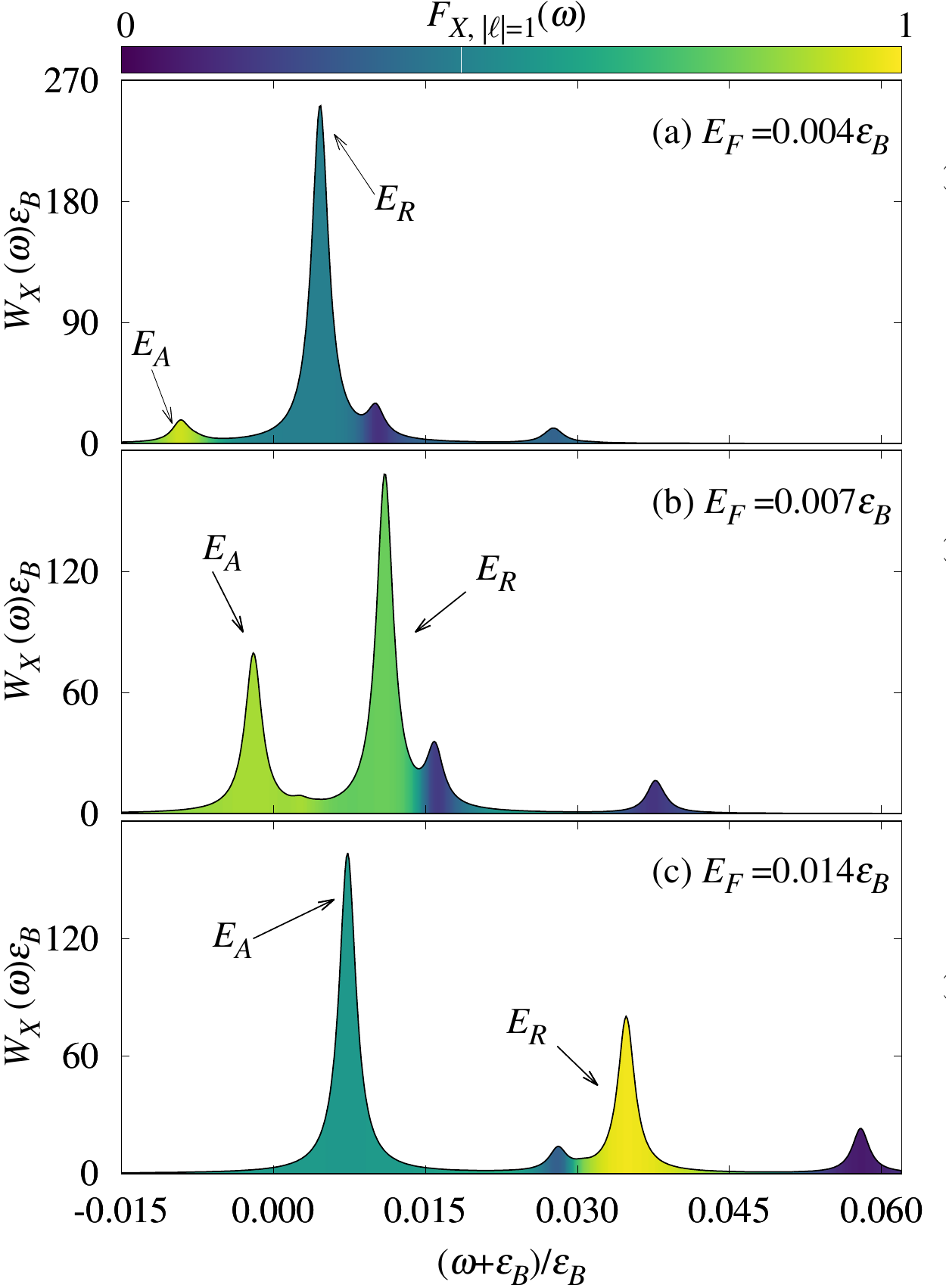}
    \caption{Exciton spectral function $W_X(\omega)$ and angular momentum character for the ICP case, in the weak coupling regime, 
    plotted for different values of $\ef$, for $m_2/m_1=0.25$, and for $N_k=10$, $N_q=4$, $N_\theta=7$. The colored area indicates the fraction of states with hole angular momentum $|\ell|=1$ in the trion-hole complex, $F_{X,|\ell|=1}(\omega)$ defined in Eq.~\eqref{eq:state-fr_ang-m1}. The line width is fixed to $\epsilon= 10^{-3}\eb$.}
    \label{fig:hole_angmom}
\end{figure}
%
\section{Hole angular momentum in the trion-hole complex}
\label{app:T3h_ang-momentum}
To gain further insight into the nature of the quasiparticle branches, we determine the probability that the hole in the trion-hole complex  has angular momentum $\ell$:
\begin{equation}
    P_{\ell}^{(n)}=  \Frac{\frac1{\area^3} \sum_{\k_1 \k_2 \q \q'} e^{i\ell(\theta_q-\theta_{q'})} \varphi_{\k_1\k_2\q}^{(n)*}  \varphi_{\k_1\k_2\q'}^{(n)} \delta_{qq'}}{\Frac{1}{\area^3} \sum_{\k_1 \k_2 \q} |\varphi_{\k_1\k_2\q}^{(n)}|^2}\; ,
\label{eq:hole-ang-mom_prob}    
\end{equation}
in a given eigenstate $n$. Here, $\theta_q$ is the angle of the majority hole momentum variable $\q=(q,\theta_q)$, see Eq.~\eqref{eq:M4_state}. 
Due to time reversal symmetry, the probability satisfies $P_{-\ell}^{(n)} = P_{\ell}^{(n)}$ and it is normalized such that $\sum_{\ell\in \mathbb{Z}} P_{\ell}^{(n)} = 1$. 
We observe that, for eigenvalues  up to the repulsive branch, $E_n \lesssim E_R$, the $|\ell|\geq2$ components have a negligible probability, so that
the hole angular momentum is either $\ell=0$ or $|\ell|=1$. Thus, in this energy interval, $P_{\ell=0}^{(n)} \simeq 1 - 2P_{\ell=1}^{(n)}$.

In order to relate this probability to frequency  $\omega$, and to focus attention on those states which are optically active, it is convenient to define the  angular-momentum-weighted exciton Green function as:
\begin{align}
\label{eq:m-weighted_GX}
    G_{X,\ell} (\omega) &= \left(\Frac{2g}{\Omega}\right)^2 \sum_n P_{\ell}^{(n)} \Frac{\left|\frac{1}{\area} \sum_{\k} \varphi_{\k}^{(n)} \right|^2 }{\omega -
E_n + i\epsilon} \; ,
\end{align}
from which we can evaluate the angular-momentum-weighted spectral function as usual:
\begin{align}
    W_{X,\ell} (\omega) &= -\Frac{1}{\pi} \Im  G_{X,\ell} (\omega) \; .
\label{eq:m-weighted_spectral-f}
\end{align}
For $\omega \lesssim E_R$, we have
\begin{align}
 W_{X}(\omega) &=\sum_{\ell\in \mathbb{Z}}W_{X,0}(\omega)\simeq W_{X,0}(\omega)+2W_{X,1}(\omega),  
\end{align}
because the $|\ell|\geq2$ hole angular momentum components are suppressed. 
As such, we may define the fraction of the spectral function with angular momentum $|\ell|=1$ as
\begin{equation}
F _{X,|\ell|=1}(\omega) \equiv \Frac{2W_{X,1}(\omega)}{W_{X}(\omega)} \; .
\label{eq:state-fr_ang-m1}
\end{equation}
When this quantity is close to zero, the hole in the trion-hole complex is predominantly $s$-wave, while a value close to one means that it is nearly all $p$-wave.
We show this in Fig.~\ref{fig:hole_angmom} by the colored area.
Note that this plot is obtained at a low resolution ($N_k=10$) because evaluating the angular momentum character $F _{X,|\ell|=1}(\omega)$ requires knowing the eigenstate in full, so we have to use a direct diagonalization routine rather than the iterative method. Nonetheless, we can still identify the attractive and repulsive branches by comparing these results with the spectral functions evaluated with higher number of points via the iterative method, which establishes which peak positions are independent of $N_k$, and may thus be identified as the attractive and repulsive branches (see Fig.~\ref{fig:M4_convergence}).

We observe that the symmetry of the  peaks that we have previously identified as the attractive and repulsive branches  evolves as a function of doping. In particular, at very low doping---see Fig.~\ref{fig:hole_angmom}(a)---as expected, the hole (and thus the trion) of the trion-hole complex in the attractive branch has  a $p$-wave symmetry, while the hole in the repulsive branch has $s$-wave symmetry. However, as $\ef$ increases, the symmetries cross over so that at larger doping---see Fig.~\ref{fig:hole_angmom}(c)---the attractive branch becomes $s$-wave and the repulsive branch $p$-wave  (see Fig.~\ref{fig:ang_momentum} in the main text). Those states associated with the 
continuum do not change symmetry and remain $s$-wave at all dopings. Note that, because attractive and repulsive peaks in Fig.~\ref{fig:hole_angmom} have constant values of the state fraction $F _{X,1}(\omega)$ within their linewidth, we can characterize this symmetry in Fig.~\ref{fig:ang_momentum} of the main text by plotting the doping dependence of the probability $P_{|\ell|=1}^{n}$ for the eigenvalue $E_n$ closest to the attractive $E_A$ and repulsive $E_R$ branches. Note that, in Fig.~\ref{fig:ang_momentum}, while the orbital character of the attractive branch is smooth, the orbital character of the repulsive branch has a kink in a small interval of $\ef$. This is due to the coupling, at finite $N_k$, between the repulsive branch and one of the continuum states, as previously explained. For this reason, we plot these data points with empty rather than filled symbols.

\begin{figure}
\includegraphics[width=0.5\textwidth]{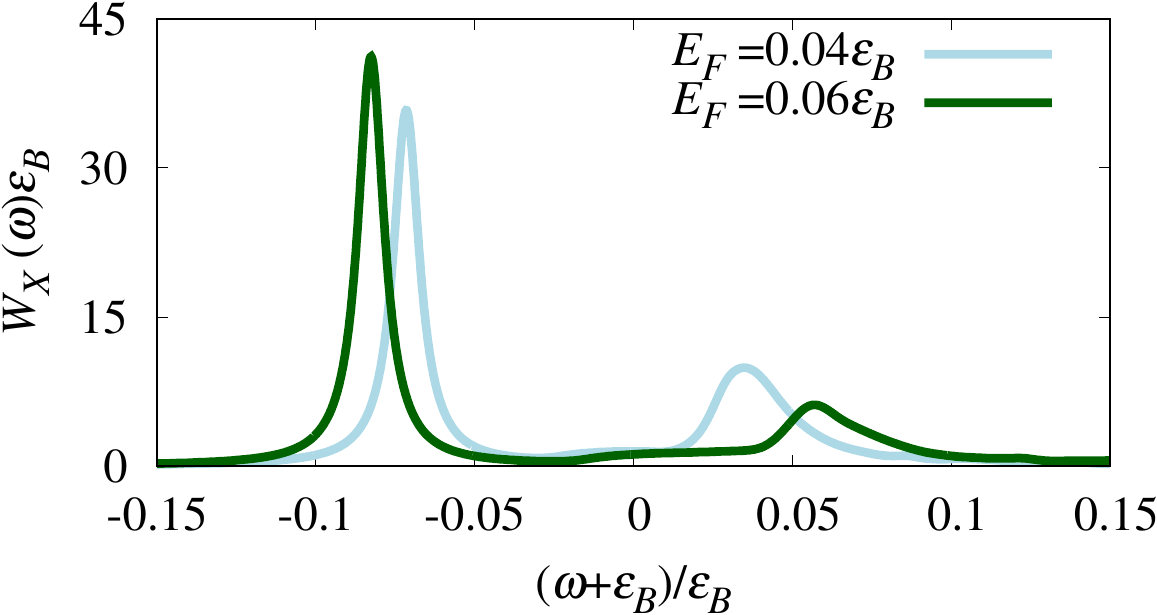}
\caption{Cross section of the exciton spectral function $W_X(\omega)$, in the 
weak coupling regime, for the DCP case sketched in Fig.~\ref{fig:schematic}(b). Parameters have been fixed to consider the specific case of a MoSe$_2$ monolayer (see text). The linewidth is $\epsilon=5\times 10^{-3}\eb$.}
\label{fig:M3-s-wave_sec}
\end{figure}
%
\section{Model of the exciton state dressing by a distinguishable Fermi sea (DCP case)}
\label{app:distinguish}
We briefly outline here the formalism describing the case where the Fermi sea dressing occurs with carriers that are distinguishable from one of the charges forming the exciton (the DCP scenario), as shown in Fig.~\ref{fig:schematic}(b),  for which the trion state is $s$-wave. Details can be found in the numerous works that study this regime, such as Refs.~\cite{Sidler_NP2016,Efimkin_PRB2017,Rana_PRL2021,Efimkin_PR2021}. For completeness, we describe briefly here the formalism that allows us to arrive at Fig.~\ref{fig:spin-swave_polaron_properties} of the main text.

The crucial distinction between the DCP and the ICP scenarios is that for distinguishable carriers the Fermi sea causes no Pauli blocking effect on the exciton. In this case, it has been shown that when the exciton binding energy is the system largest scale, i.e., much larger than the Fermi energy and the $s$-wave trion binding energy, the system is well described by approximating the exciton as a tightly bound bosonic particle~\cite{Efimkin_PR2021}.
Adopting this widely-used tightly-bound exciton approximation, we consider the following Hamiltonian:
\begin{subequations}
\label{eq:Hamiltonian_swave}
\begin{align}
\label{eq:Hamiltonian_terms_swave}
  \hat{H} &= \hat{H}_0 + \hat{H}_{Xe} +
  \hat{H}_{XC}\\
\label{eq:H_0_swave}
  \hat{H}_0 &=\sum_{\k} \epsilon_{\k}  \hat{c}^\dag_{\k} \hat{c}^{}_{\k}+ \sum_{\q} \omega_{X\q} \hat{x}_{\q}^{\dag} \hat{x}_{\q}^{}
  + \sum_{\q} \nu_{\q} \hat{a}_{\q}^{\dag} \hat{a}_{\q}^{} \\
\label{eq:H_Xe_swave}
  \hat{H}_{Xe} &= -\frac{u}{\area}\sum_{\ve{k}\ve{k'}\ve{q}}
  \hat{x}^\dag_{\ve{k}} \hat{c}^\dag_{\ve{k'}}
  \hat{c}^{}_{\ve{k'}+\ve{q}}
  \hat{x}^{}_{\ve{k}-\ve{q}}\\
  \hat{H}_{XC} &= g \sum_{\q}
  \left(\hat{x}^\dag_{\q} \hat{a}_{\q}^{} + \text{h.c.}\right) \; .
\label{eq:H_XC_swave}
\end{align}
\end{subequations}
The electrons of the Fermi sea are described by the operators $\hat{c}_{\k}^{}$ and have a dispersion $\epsilon_{\k} = \k^2/2m_e$. The excitons are described by the bosonic operators $\hat{x}_{\q}^{}$ and have a dispersion $\omega_{X\q}=-\eb+\q^2/2m_X$, where $m_X=m_e+m_h$ is the exciton mass. We consider here the  specific parameters for a MoSe$_2$ monolayer, in particular we fix the exciton binding energy $\eb=500$~meV~\cite{Sidler_NP2016} and electron $m_e=0.55 m_0$ and hole $m_h=0.59m_0$ effective masses~\cite{Berkelbach_PRB2013, Rasmussen_JPCC2015}, where $m_0$ is the free electron mass.
The electron-exciton interaction can be safely approximated as contact~\cite{Sidler_NP2016,Efimkin_PR2021} with strength $u$. As seen in the main text, this needs to be renormalized, which  can be done by introducing the trion binding energy $\varepsilon_T=|E_T| - \eb$. In the specific approximation considered here, the trion becomes a two-body problem of exciton-electron pairing. Thus the analog of Eq.~\eqref{eq:v_renormalization} here is to write:
\begin{equation}
    \Frac{1}{u} = \Frac{1}{\area} \sum_\k^{\Lambda}
    \frac{1}{\varepsilon_T+\omega_{X\k}+\epsilon_{\k} } \; .
\label{eq:v_renormalization_swave}
\end{equation}
We consider the specific value of the trion binding energy $\varepsilon_T = 25$~meV for MoSe$_2$ monolayers~\cite{Sidler_NP2016}.
Finally, the cavity photons are described by the operators $\hat{a}_{\q}^{}$ and dispersion $\nu_{\q} = \delta -\eb + \q^2/2m_C$, where $\delta$ is the photon-exciton detuning.
The contact matter-light coupling has a strength $g$; because the exciton is structureless, the matter-light coupling $g$ does not require any renormalization, and the polariton Rabi splitting is given by $\Omega=2g$.

In order to evaluate the system spectral response, we consider the following zero-momentum polaron ansatz, which considers an exciton, a photon state, and the single electron-hole dressing of the Fermi sea generated by the presence of the exciton, which describes the trion-hole~\cite{Sidler_NP2016}:
\begin{equation}
  |\widetilde{P}_3\rangle=\left(\varphi \hat{x}^{\dag}_{\0}+\alpha\hat{a}^{\dag}_{\0}+\sum_{\k\q} \Frac{\varphi_{\k\q}}{\area} \hat{x}^{\dag}_{\q-\k}\hat{c}^{\dag}_{\k}\hat{c}_{\q} \right)\ket{FS} \; .
\label{eq:M3_state}
\end{equation}
As before, momenta $\k$ are $k>\kF$ and $\q$ are $q<\kF$. The eigenvalue equations obtained by minimising  $\bra{\widetilde{P}_3}(\hat{H}-E)\ket{\widetilde{P}_3}$ with respect to $\alpha^*$, $\varphi^*$, and  $\varphi_{\k\q}^*$ are:
%
\begin{align*}
    E\alpha &= \nu_{\0} \alpha - g\varphi \\
    E\varphi &= \omega_{X\0} \varphi - g\alpha - \Frac{u}{\area^2} \sum_{\k\q} \varphi_{\k\q} \\
    E\varphi_{\k\q} &= E_{X\k\q}\varphi_{\k\q} - \frac{u}{\area}\sum_{\k'}\varphi_{\k'\q} + \frac{u}{\area} \sum_{\q'}\varphi_{\k\q'} - u\varphi\; ,
\end{align*}
%
with $E_{X\k\q}=\omega_{X\q-\k}+\epsilon_\k-\epsilon_\q$. As in Eqs.~\eqref{eq:M4_eigenvalue}, the trion-hole term $\varphi_{\k\q}$ does not couple directly to the photon term $\alpha$, rather it couples indirectly via the exciton amplitude $\varphi$.

In order to obtain 
Fig.~\ref{fig:spin-swave_polaron_properties} of the main text, we solve the eigenvalue equations in the weak-coupling regime 
by discretizing the momenta on a grid as described in Appendix~\ref{app:mom_grid},
renormalize the interaction strength $u$ via Eq.~\eqref{eq:v_renormalization_swave}, and evaluate the exciton spectral function~\eqref{eq:spectral-f_X} from exciton Green's function
\begin{equation}
    G_X^{(0)}(\omega) = \sum_n \Frac{|\varphi^{(n)}|^2}{\omega-E_n+i\epsilon} \; .
\end{equation}
We then increase 
the grid number of points until convergence is reached.
Because the exciton is tightly bound, there is no need to renormalize the exciton Green's function. 

In Fig.~\ref{fig:M3-s-wave_sec} we show the resulting exciton spectral function at two different densities. This figure is obtained without applying the Gaussian convolution that was used in the ICP case, because, for this model, the numerics allows us to reach large enough $N_k$ to get convergence for the entire spectrum, including the 
continuum.


%

\end{document}